\pdfoutput=1
\documentclass[aps,showpacs,twocolumn,twoside,pre,superscriptaddress]{revtex4-1}
\usepackage{epsfig,amsmath,amssymb}
\usepackage{graphicx}
\usepackage{array}
\usepackage{xcolor}

\def\I#1{\!#1\!}

\def\ket#1{|#1\rangle}
\def\bra#1{\langle#1|}
\def\scal#1#2{\langle#1|#2\rangle}
\def\matr#1#2#3{\langle#1|#2|#3\rangle}

\def\ave#1{\langle#1\rangle}
\def\ove#1{\overline{#1}}

\def\={\!=\!}
\def\>{\!>\!}
\def\<{\!<\!}
\def\-{\!-\!}
\def\+{\!+\!}

\def\uvo#1{\lq\lq #1\rq\rq}

\begin{document}

\title{Superradiance in finite quantum systems randomly coupled to continuum}

\author{Pavel Str\'ansk\'y}
\email{stransky@ipnp.troja.mff.cuni.cz}
\affiliation{Institute of Particle and Nuclear Physics, Faculty of Mathematics and Physics, Charles University, V Hole\v{s}ovi\v{c}k\'ach 2, 18000 Prague, Czech Republic} 
\author{Pavel Cejnar}
\email{cejnar@ipnp.troja.mff.cuni.cz}
\affiliation{Institute of Particle and Nuclear Physics, Faculty of Mathematics and Physics, Charles University, V Hole\v{s}ovi\v{c}k\'ach 2, 18000 Prague, Czech Republic} 

\date{\today}

\begin{abstract}
We study the effect of superradiance in open quantum systems, i.e., the separation of short- and long-living eigenstates when a certain subspace of states in the Hilbert space acquires an increasing decay width. 
We use several Hamiltonian forms of the initial closed system and generate their coupling to continuum by means of the random matrix theory.
We average the results over a large number of statistical realizations of an effective non-Hermitian Hamiltonian and relate robust features of the superradiance process to the distribution of its exceptional points.
We show that the superradiance effect is enhanced if the initial system is at the point of quantum criticality.
\end{abstract}

\maketitle

\section{Introduction}
\label{INTR}

Imagine a classical system that randomly decays with a rate $\gamma$ whenever it passes a certain decay zone in the phase space.
For time $\Delta t$ spent in the zone, $e^{-\gamma\Delta t}$ is the survival probability and ${(1\I{-}e^{-\gamma\Delta t})}$ is the probability for the system to \uvo{disappear}. 
As the phase space is assumed to describe the composite system and not the constituents into which it disintegrates, any phase-space probability density gradually looses its normalization to unity.
The accessibility of the decay zone depends on energy $E$ determined by the system's Hamiltonian.
Minimal and maximal Hamiltonian values $E_{\rm min}$ and $E_{\rm max}$ within the zone (which is assumed to be compact) determine an energy window in which the system gets unstable. 
Assuming ergodicity of classical motions, all states within the above interval will be characterized by  decay rates $\Gamma(E)$, which depend on details of dynamics but are proportional to $\gamma$ for any fixed energy.

An analogous setup in the {\em quantum\/} world has significantly different consequences.
Consider a simplified model of a bound quantum system with $d$-dimensional Hilbert space ${\cal H}$ of state vectors.
The quantum decay zone is defined as a certain $n$-dimensional subspace ${{\cal H}_{\rm D}\subset{\cal H}}$, which couples uniformly to an infinite environment with continuous spectrum.
Consequently, all states within ${\cal H}_{\rm D}$ have the same decay width $\gamma\I{=}\hbar/(2\tau)$, where $\tau$ stands for the mean lifetime (here we define \uvo{energy width} consistently with relation ${\gamma\tau=\frac{1}{2}\hbar}$ instead of a more common definition without factor $\frac{1}{2}$).
Since any eigenstate of an independently chosen Hamiltonian has generally a non-vanishing overlap with ${\cal H}_{\rm D}$, the whole system under the Hamiltonian-induced evolution becomes unstable.
One can apply a simple description based on the Feshbach theory \cite{Fes,Moi11}, in which the open system is attributed by an effective non-Hermitian Hamiltonian $\hat{H}^{(\gamma)}\I{=}{\hat{H}^{(0)}\I{-}i\gamma\hat{P}_{\rm D}}$, with $\hat{H}^{(0)}$ denoting Hamiltonian of the unperturbed closed system and $\hat{P}_{\rm D}$ a projector to ${\cal H}_{\rm D}$.
The eigenstates $\ket{\kappa^{(\gamma)}}$ of $\hat{H}^{(\gamma)}$ for any finite value of $\gamma$ have generally non-zero decay widths $\Gamma^{(\gamma)}_{\kappa}\I{=}\hbar/(2\tau^{(\gamma)}_{\kappa})$. 
If $\gamma$ grows infinitesimally from zero, $\Gamma^{(\gamma)}_{\kappa}$ of all eigenstates increase proportionally.
However, as the width $\gamma$ grows to higher values, the set of eigenstates eventually splits into two groups:
The first one contains $n$ states whose widths keep growing as $\Gamma^{(\gamma)}_{\kappa}\I{\approx}\gamma$.
In contrast, the widths of remaining ${(d-n)}$ states in the second group overcome the initially increasing trend and return back to zero, ${\Gamma^{(\gamma)}_{\kappa}\to 0}$, with ${\gamma\to\infty}$.
For a sufficiently high value of $\gamma$, the original real spectrum 
is reorganized so that it consists of $n$ very wide and ${(n\I{-}d)}$ very narrow states with modified but not too distant real energies.
An example of such behavior is shown in Fig.\,\ref{trajekt} below.

The above-described phenomenon was probably for the first time pointed out in Ref.\,\cite{Kle85}.
Its mechanism was soon related \cite{Sok88,Sok92} to so-called Dicke superradiance \cite{Dic54}, in which mutual coupling mediated by a common electromagnetic field in a dense ensemble of atoms leads to a collective enhancement and time squeeze of the spontaneous radiation emitted from the ensemble.
In the case of an open system, the role of the mediating field is played by mutual coupling of the unperturbed eigenstates by the decay-inducing part of the full Hamiltonian, and an analog of the superradiant burst is the creation of the group of very short-living (super-radiant) states on a background of long-living (sub-radiant) ones.
To distinguish the latter effect from the original notion of superradiance, we call it {\em non-Hermitian superradiance\/} (NHSR), emphasizing the non-Hermiticity of the model Hamiltonian which captures the coupling of its eigenstates to the continuum. 

The NHSR has notable implications in complex many-body systems, such as nuclei, baryon excitations, atoms, atomic clusters and molecules, open quantum system with gain and loss etc. \cite{Rot91,Vol04,Aue11,Cel11,Liu14,Ele14,Rot15,Gre15,Ele17a,Ele17b,Jos18}. 
Atomic nuclei, in particular, show neat examples of narrow quasi-stationary states (such as neutron or proton resonances) coexisting with much broader structures (various kinds of doorway states or giant resonances), and the above toy model of resonance trapping provides an elementary background for their description.
It is relevant also in various artificial quantum systems realizable with the aid of recent laboratory quantum simulators \cite{Liu14,Rot15,Gre15,Jos18}. 
For detailed reviews with additional references see Refs.\,\cite{Aue11,Ele14,Rot15}.

Essential insight into the mechanism of the NHSR follows from the mathematics of so-called exceptional points (EPs) \cite{Kat66,Zir83}. 
These represent degeneracies of a Hamiltonian $\hat{H}^{(\lambda)}$ with a real discrete spectrum depending on parameter $\lambda$ in complex-extended domain ${\lambda\in\mathbb{C}}$, where $\hat{H}^{(\lambda)}$ is non-Hermitian and its eigenvalues complex.
Convergence of EPs to a point $\lambda_{\rm c}$ on the real parameter axis with asymptotically increasing size of the system was shown to trigger quantum phase transitions (QPTs) \cite{Hei89,Cej07,Sin17,Str18}.
In the setup of an open quantum system, the EPs are responsible for the NHSR-type redistribution of complex eigenvalues of the Hamiltonian with a running decay rate $\gamma$ \cite{Ele14,Hei98,Jun99}.

The role of EPs in the NHSR process was studied in Refs.\,\cite{Ele14,Hei98,Jun99}.
In the present paper we extend those studies in several directions:
First, we generate the decay-inducing part of the Hamiltonian by means of the random matrix theory.
Statistical averaging over an ensemble of Hamiltonian realizations yields robust results, washing out any particular correlation between the initial eigenbasis and the set of open states.
Second, we investigate the dependence of the superradiance process on the form of the initial Hamiltonian.
In particular, we connect the two above-mentioned roles of EPs showing that if the initial Hamiltonian is at the quantum critical point, the system exhibits sharper NHSR dynamics. 

The outline of the paper is as follows:
We first describe our statistical model to study the NHSR phenomenon (Sec.\,\ref{HAMI}), demonstrate on it a general effect of EPs (Sec.\,\ref{BIDEC}), and derive some overall properties of the complex spectrum (Sec.\,\ref{GLOPR}). 
We further analyze special properties of the NHSR arising from criticality of the initial Hamiltonian, in particular from its association with the first- and second-order QPT (Sec.\,\ref{CRIH}). 
At last, we summarize the results (Sec.\,\ref{SUMA}).

\section{Model Hamiltonian}
\label{HAMI}

\begin{figure}[t!]
\includegraphics[width=\linewidth]{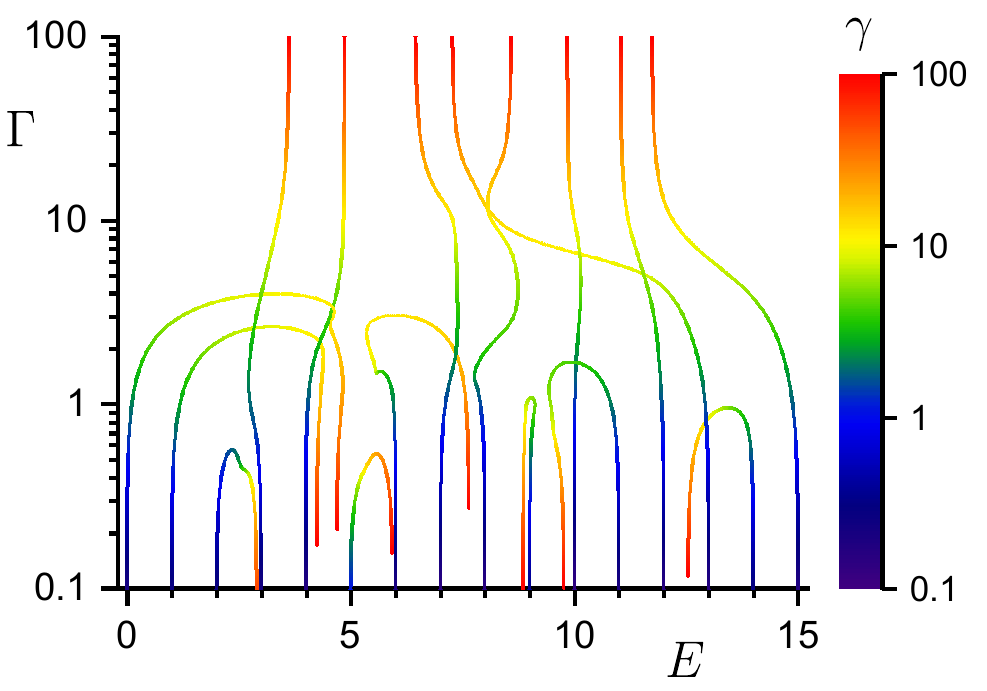}
\caption{
Trajectories of complex eigenvalues ${\cal E}_{\kappa}^{(\lambda)}\I{=}E_{\kappa}^{(\lambda)}\I{-}i\Gamma_{\kappa}^{(\lambda)}$ of Hamiltonian \eqref{H} with running parameter ${\gamma=-{\rm Im}\lambda}$ (its value is expressed by color) and ${\epsilon={\rm Re}\lambda=0}$ for ${d=16}$ and ${n=8}$.
The initial Hamiltonian ${\hat{H}^{(0)}=\hat{H}^{(0)}_{\rm HO}}$ has equidistant spectrum with unit spacing, while $n$ decaying states $\ket{\phi_l}$ are samples from random GOE eigenvectors (see Sec.\,\ref{HAMI}).
}
\label{trajekt}
\end{figure}   

In this section we introduce a simple model of NHSR used in our analysis.
The model works in a finite, $d$-dimensional Hilbert space ${\cal H}$, in which $n\in[1,d-1]$ orthogonal states are supposed to be equally coupled to continuum.
An example of the superradiant separation of short- and long-living states for this model is shown in Fig.\,\ref{trajekt}.
Everywhere we set ${\hbar=1}$.

At first we consider a general Hamiltonian of the form
\begin{equation}
\hat{H}^{\left(\{\lambda_l\}\right)}=\underbrace{\sum_{k=1}^{d}E_{k}^{(0)}\ket{k}\bra{k}}_{\hat{H}^{(0)}}+\sum_{l=1}^{n}\underbrace{(\epsilon_l-i\gamma_l)}_{\lambda_l}\ket{\phi_l}\bra{\phi_l}
\label{Hagen}
\,,
\end{equation}
where the first term $\hat{H}^{(0)}$ (the unperturbed initial Hamiltonian) describes a closed system with energies ${E_{k}^{(0)}\in\mathbb{R}}$ and orthonormal eigenvectors $\ket{k}$, while the second term (decay-inducing Hamiltonian) defines $n$ decaying states given by orthonormal vectors $\ket{\phi_l}$  with complex energies ${\lambda_l\in{\mathbb C}}$, each composed of a real part ${\epsilon_l\in\mathbb{R}}$ and a non-negative decay width ${\gamma_l\in\mathbb{R}}$.
The orthonormality conditions read as ${\scal{k_1}{k_2}=\delta_{k_1k_2}}$ and ${\scal{\phi_{l_1}}{\phi_{l_2}}=\delta_{l_1l_2}}$, where both vector sets $\{\ket{k}\}_{k=1}^d$ and $\{\ket{\phi_l}\}_{l=1}^n$ are supposed to be incompatible.

To illustrate physical meaning of the schematic Hamiltonian \eqref{Hagen}, we assume a system of $N$ fermions in a finite Hilbert space generated by $d_{\rm sp}$ single-particle states in a mean-field potential well.
Out of these states, $(d_{\rm sp}\I{-}n_{\rm sp})$  are stable, bounded inside the well below the continuum threshold energy, while the remaining $n_{\rm sp}$ states are quasi-stable, located above the continuum threshold but confined with large lifetimes in the potential well region due to a barrier.
For $N$ fermions, the total Hilbert space of dimension $d\I{=}\binom{d_{\rm sp}}{n_{\rm sp}}$ is spanned by $\binom{d_{\rm sp}-n_{\rm sp}}{N}$ stable and $n\I{=}\binom{d_{\rm sp}}{N}\I{-}\binom{d_{\rm sp}-n_{\rm sp}}{N}$ unstable mean-field configurations.
The unstable subspace is generated by $N$-body basis vectors $\{\ket{\phi_l}\}_{l=1}^{n}$ in which at least one fermion is in the unstable single-particle state, the stable subspace is generated by the remaining vectors $\{\ket{\phi_l}\}_{l=n+1}^{d}$. 
Now we assume that all fermions are subject to some residual two-body interactions acting between all $d_{\rm sp}$ single-particle states.
The real parts of the unperturbed energies and the residual interaction between individual mean-field states (its matrix elements can be calculated from wave functions restricted to the domain inside the well) jointly form a Hermitian Hamiltonian $\hat{H}^{(0)}$  which approximately describes the system without tunneling.
The eigenbasis $\{\ket{k}\}_{k=1}^{d}$ of $\hat{H}^{(0)}$ is rotated relative to the mean-field basis $\{\ket{\phi_l}\}_{l=1}^{d}$.
The decay describing part of the Hamiltonian consists of projectors to original unstable states $\{\ket{\phi_l}\}_{l=1}^{n}$ with complex coefficients expressing decay widths $\gamma_l$ and real-energy corrections $\epsilon_l$ depending on the shape of the confining potential barrier.
This results in a non-Hermitian Hamiltonian of the form \eqref{Hagen}.

In the following, we will use a simplified version of this Hamiltonian, namely
\begin{equation}
\hat{H}^{(\lambda)}=\hat{H}^{(0)}+\underbrace{(\epsilon-i\gamma)}_{\lambda}\underbrace{\sum_{l=1}^{n}\ket{\phi_l}\bra{\phi_l}}_{\hat{P}_{\rm D}}
\label{H}
\,.
\end{equation}
Here, the decay-inducing part of the Hamiltonian $\lambda\hat{P}_{\rm D}=\hat{H}_{{\rm D}}^{(\lambda)}$ is determined by a single projection operator $\hat{P}_{\rm D}$ and all the decaying states, which form a subspace ${{\cal H}_{\rm D}=\hat{P}_{\rm D}{\cal H}\subset{\cal H}}$, have the same complex energy ${\epsilon-i\gamma=\lambda}$.
The imaginary component $\gamma$ expresses the decay width of states in ${\cal H}_{\rm D}$ and the real component $\epsilon$ represents their energy shift. 
We stress that though the simplified Hamiltonian \eqref{H} restricts direct applicability of our model to specific systems, it captures essential features of the more complex Hamiltonian \eqref{Hagen} while considerably reducing the number of free parameters.  

The decay-inducing term of Hamiltonian \eqref{H} can be cast in the form ${\hat{H}_{{\rm D}}^{(\lambda)}=\lambda\hat{P}_{\rm D}+0\hat{P}_{\bot}}$ (with $\hat{P}_{\bot}=\hat{\mathbb{I}}-\hat{P}_{\rm D}$) with two eigenvalues $\lambda$ and 0.
The eigenspace associated with eigenvalue $\lambda$ coincides with the decaying subspace ${\cal H}_{\rm D}$ of dimension $n$, while the eigenspace with eigenvalue $0$ is the subspace ${{\cal H}_{\bot}=\hat{P}_{\bot}{\cal H}}$ with dimension ${n_{\bot}=d-n}$. 
Note that the cases ${n=0}$ with ${\hat{H}_{{\rm D}}^{(\lambda)}=0}$ and ${n=d}$ with ${\hat{H}_{{\rm D}}^{(\lambda)}=\lambda\hat{\mathbb{I}}}$ are both trivial and we exclude them.

In our analysis, the unperturbed component $\hat{H}^{(0)}$ of the total Hamiltonian \eqref{H} is associated with the Lipkin-Meshkov-Glick model \cite{Lip65}. 
It is build from quasispin operators ${(\hat{J}_1,\hat{J}_2,\hat{J}_3)}$ satisfying the standard angular-momentum commutation relations.
The conserved quantity $\hat{J}^2$ is fixed at the value $ j(j+1)$, so that the Hilbert space ${\cal H}$ spanned by eigenvectors of $\hat{J}_3$ has dimension ${d=2j+1}$.
To generate the spectrum of energies $E_k^{(0)}$, we consider the following three alternative forms:
\begin{eqnarray}
\hat{H}^{(0)}&=&\hat{H}^{(0)}_{\rm HO}\ =s_{0}(j)\,\hat{J}_3+a_0(j)
\,,\label{H0ho}\\
&=&\hat{H}^{(0)}_{\rm PT1}=s_{1}(j)\left(\hat{J}_3-\frac{3}{j}\hat{J}_1^2\right)+a_1(j)
\,,\label{H0pt1}\\
&=&\hat{H}^{(0)}_{\rm PT2}=s_{2}(j)\left(\hat{J}_3-\frac{1}{2j}\hat{J}_1^2\right)+a_2(j)
\,,\label{H0pt2}
\end{eqnarray}
where $s_{i}(j)$ and $a_{i}(j)$ with $i\I{=}0,1,2$ are dimension-dependent scaling and shift constants ensuring invariant bounds $E_1^{(0)}\I{=}0$ and $E_d^{(0)}\I{=}d$ of the unperturbed spectrum and therefore a unit average spacing between neighboring energy levels in all cases \eqref{H0ho}--\eqref{H0pt2}.
We introduce cumulants expressing global properties of the unperturbed spectrum, in particular the average $\overline{E}^{(0)}=\frac{1}{d}\sum_{k=1}^dE_k^{(0)}$, which represents a central energy of the spectrum, the variance $\overline{\Delta^2 E}^{(0)}=\frac{1}{d}\sum_{k=1}^d\bigl[E_k^{(0)}\I{-}\overline{E}^{(0)}\bigr]^2$, which characterizes a squared spread of the spectrum, and the skewness $\overline{\Delta^3 E}^{(0)}=\frac{1}{d}\sum_{k=1}^d\bigl[E_k^{(0)}\I{-}\overline{E}^{(0)}\bigr]^3$, which quantifies an asymmetry of the spectrum with respect to $\overline{E}^{(0)}$.
We note that all results below are independent of an overall shift of the spectrum, while a change of the average spacing between levels (spread of the spectrum) can be compensated by an appropriate rescaling of parameter $\lambda$.  

\begin{figure}[t!]
\includegraphics[width=0.85\linewidth]{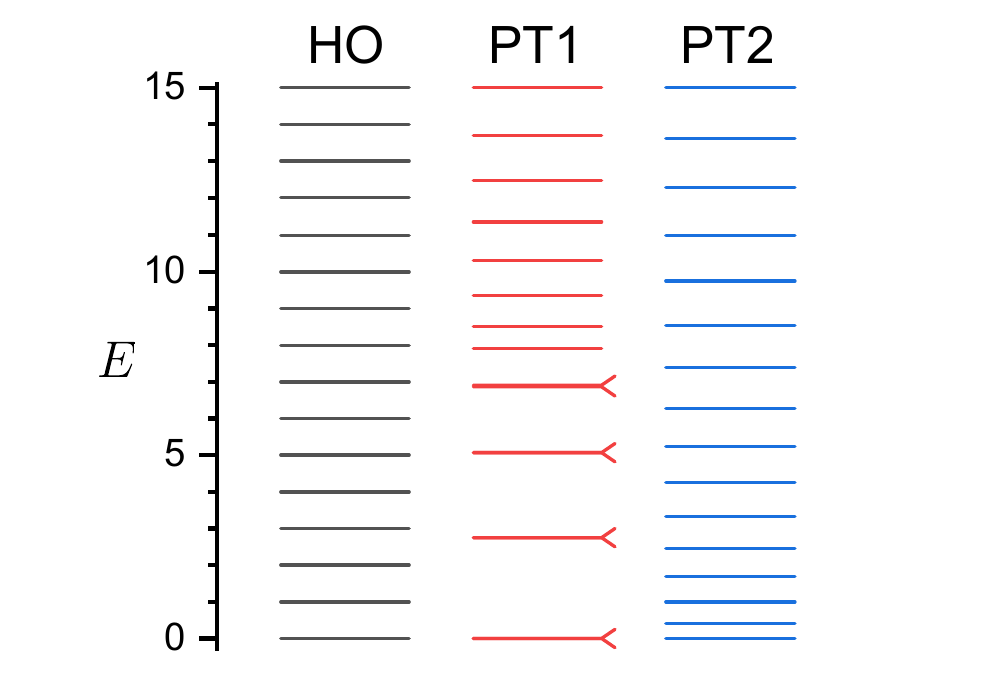}
\caption{
Energy spectra of the initial Hamiltonian $\hat{H}^{(0)}$ from Eqs.\,\eqref{H0ho}--\eqref{H0pt2} for ${d=16}$.
Marks \uvo{$<$} indicate doublets of levels.
}
\label{spek}
\end{figure}   

All Hamiltonians \eqref{H0ho}--\eqref{H0pt2} can be written, using the Holstein-Primakoff transformation \cite{Hol40}, in terms of a single pair ${(\hat{q},\hat{p})}$ of conjugate coordinate and momentum operators satisfying in the classical limit a constraint ${q^2+p^2\leq 2}$ (for a general discussion of this mapping see, e.g., \cite{Str19}).
The Hamiltonian $\hat{H}^{(0)}_{\rm HO}$ (the abbreviation standing for \uvo{harmonic oscillator}) represents an equidistant spectrum composed of $d$ levels with spacing ${E^{(0)}_{k+1}-E^{(0)}_{k}=1}$ associated with a potential ${V(q)\propto q^2+{\rm const}}$.
Hamiltonians $\hat{H}^{(0)}_{\rm PT1}$ and $\hat{H}^{(0)}_{\rm PT2}$ (\uvo{phase transitional} of the first and second kind) describe representative $d$-dimensional spectra of two types of quantum critical systems:
$\hat{H}^{(0)}_{\rm PT1}$ stands for the first-order QPT Hamiltonian with energy spectrum consisting of parity doublets of levels in a degenerate symmetric double-well potential ${V(q)\propto 3q^4-5q^2+{\rm const}}$. 
The spacing between the low-lying states forming the doublets decreases with dimension as ${E^{(0)}_{k+1}-E^{(0)}_{k}\propto e^{-c_kd}}$ (where $c_k$ is a constant), while spacing between states outside the doubles remains of order ${E^{(0)}_{k+2}-E^{(0)}_{k+1}\sim O(1)}$.
$\hat{H}^{(0)}_{\rm PT2}$ represents a second-order QPT Hamiltonian associated with a quartic potential ${V(q)\propto q^4+{\rm const}}$. 
Its spectrum exhibits a typical power-law cumulation of low-energy levels according to ${E^{(0)}_{k+1}-E^{(0)}_{k}\propto\left(k/d\right)^{1/3}}$.
Energy levels of all the three initial Hamiltonians are for a moderate dimension shown in Fig.\,\ref{spek}. 
More details concerning the critical Hamiltonians can be found in Ref.\,\cite{Str18}. 

The decay-inducing component of Hamiltonian \eqref{H} is represented not by a single fixed operator $\hat{H}_{{\rm D}}^{(\lambda)}$, but by a suitable {\em statistical ensemble\/} of its possible (in some sense equivalent) realizations. 
The results are obtained by averaging over a large number $N_{\rm R}$ (of order from $10^1$ to $10^4$) of samples from this ensemble.
The orthonormal vectors $\{\ket{\phi_l}\}_{l=1}^n$ and $\{\ket{\phi_{\bot l'}}\}_{l'=1}^{n_{\bot}}$ forming in each realization the bases of subspaces ${\cal H}_{\rm D}$ and ${\cal H}_{\bot}$, respectively, result from a random orthogonal transformation of the original eigenbasis $\{\ket{k}\}_{k=1}^d$.

To achieve a completely unbiased (isotropic) generation of these bases, we use the eigenstate components of matrices from the Gaussian orthogonal ensemble (GOE), which has the equivalence of bases in its definition \cite{Meh04}.
In particular, for each realization we perform the following steps: (i) we generate a random $d$-dimensional real symmetric matrix ${\hat{H}_{\rm GOE}\in{\rm GOE}}$, i.e., a matrix with independent elements taken from zero-mean Gaussian distributions with ${\sigma^2=2}$ or 1 for diagonal or off-diagonal elements, respectively, (ii) we find an orthogonal matrix $\hat{O}$ transforming $\hat{H}_{\rm GOE}$ to the diagonal form ${\hat{D}=\hat{O}^{\rm T}\hat{H}_{\rm GOE}\hat{O}}$, where ${\hat{O}^{\rm T}=\hat{O}^{\dag}=\hat{O}^{-1}}$  stands for the transpose of $\hat{O}$, (iii) we randomly chose $n$ columns of $\hat{O}$ and associate them with the decaying states $\{\ket{\phi_l}\}_{l=1}^n$.
This procedure is repeated until a large number $N_{\rm R}$ of $\hat{H}_{\rm GOE}$ realizations yields a robust estimate of the analyzed quantities.
The averaging over the random-matrix ensemble will be further denoted by angular brackets $\ave{\bullet}$, which is in contrast to the bar symbol $\overline{\bullet}$ introduced above for the \uvo{non-statistical} averages over the energy levels.
Note that orthogonality of the transformation $\hat{O}$, which is inherent in the foundations of GOE, guarantees that matrix elements ${\matr{k_1}{\hat{P}_{\rm D}}{k_2}=\sum_{l=1}^n\scal{k_1}{\phi_l}\scal{\phi_{l}}{k_2}=\sum_{l=1}^nO_{k_1l}O_{k_2l}}$ are real, so that we can write: ${{\rm Im}{\matr{k_1}{\hat{H}(\lambda)}{k_2}}\propto{\rm Im}\lambda=-\gamma}$.

\section{Bifurcation of decay widths}
\label{BIDEC}

\begin{figure}[t!]
\includegraphics[width=\linewidth]{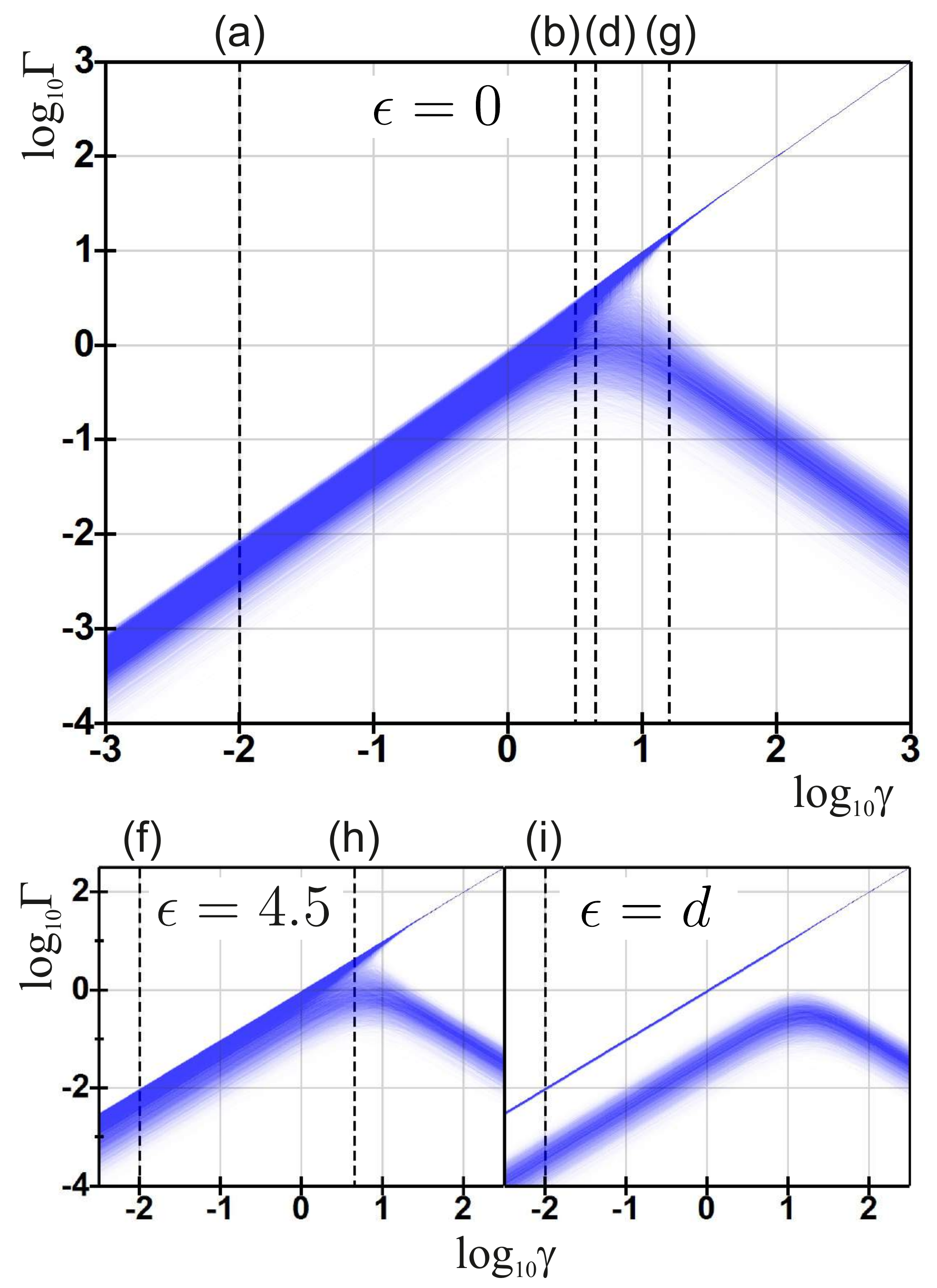}
\caption{
The evolution of decay widths $\Gamma_{\kappa}^{(\lambda)}$ with increasing $\gamma$ for Hamiltonian \eqref{H} with ${\hat{H}^{(0)}=\hat{H}^{(0)}_{\rm HO}}$ and ${d=2n=16}$ averaged over a large number $N_{\rm R}$ (=$1200$ in upper panel and 800 in lower panels) GOE realizations of the decaying subspace.
The upper panel corresponds to ${\epsilon=0}$, the lower panels to ${\epsilon=4.5}$ and ${\epsilon=16}$.
Each picture consists of $d\cdot N_{\rm R}$ tiny curves corresponding to the evolution of decay widths of all levels in all realizations, so higher color intensity implies higher probability of the corresponding value.
Samples of these distributions at values of $\gamma$ corresponding to the dashed vertical lines are shown in the respective panels of Fig.\,\ref{widths1}.
The observed behavior can be compared with decay width evolutions for a ${d=2}$ model in Fig.\,\ref{2D}.  
}
\label{bands}
\end{figure}   

\begin{figure}[t!]
\includegraphics[width=\linewidth]{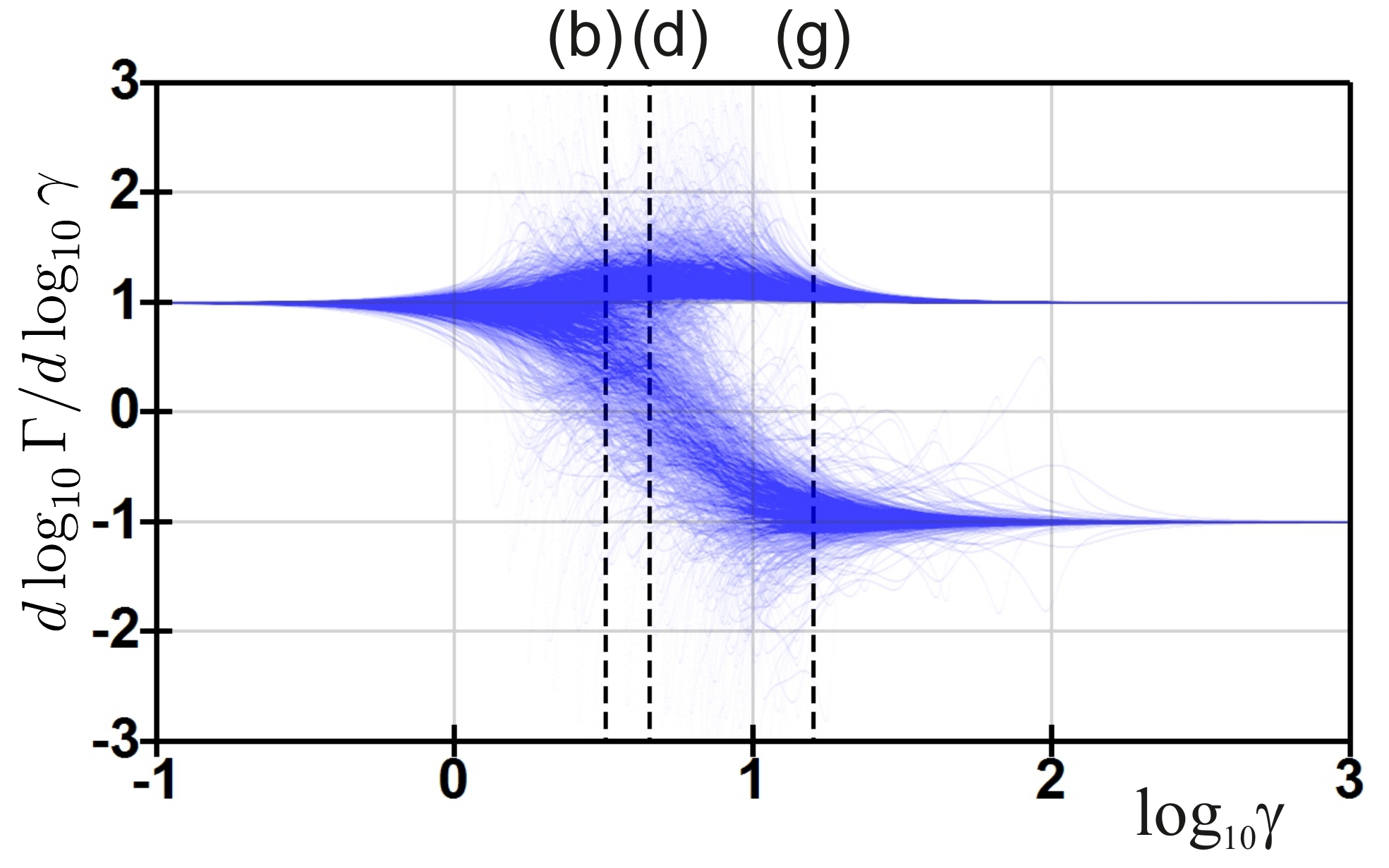}
\caption{
The evolution of log-log slopes $\frac{d}{d\log\gamma}\log\Gamma_{\kappa}^{(\lambda)}$ of decay widths from the main ($\epsilon\I{=}0$) panel of Fig.\,\ref{bands} with increasing $\gamma$ for $N_{\rm R}\I{=}100$ realizations of the decaying subspace (${d=2n=16}$).
The picture consists of $d\cdot N_{\rm R}$ tiny curves displaying the evolution of slopes for all levels in all realizations.
}
\label{slopes}
\end{figure}   

In the following, we focus on the dependencies of the eigensolutions of the Hamiltonian \eqref{H} on parameter ${\lambda=\epsilon-i\gamma}$. 
The Hamiltonian is a complex matrix symmetric under transposition (but not under the full Hermitian conjugation) with $d$ generally complex eigenvalues ${{\cal E}_{\kappa}^{(\lambda)}=E_{\kappa}^{(\lambda)}-i\Gamma_{\kappa}^{(\lambda)}}$. 
The real parts $E_{\kappa}^{(\lambda)}$ stand for energies and the imaginary parts ${\Gamma_{\kappa}^{(\lambda)}\geq 0}$ represent decay widths of individual eigenstates enumerated by integer ${\kappa=1,2,\dots,d}$.
Because of non-Hermiticity, the eigenstates of $\hat{H}^{(\lambda)}$ must be distinguished to right eigenstates satisfying ${\hat{H}^{(\lambda)}\ket{\kappa_{\rm R}^{(\lambda)}}={\cal E}_{\kappa}^{(\lambda)}\ket{\kappa_{\rm R}^{(\lambda)}}}$, and left eigenstates satisfying  $\bra{\kappa_{\rm L}^{(\lambda)}}{\hat{H}^{(\lambda)}=\bra{\kappa_{\rm L}^{(\lambda)}}{\cal E}_{\kappa}^{(\lambda)}}$.
From the transposition symmetry of $\hat{H}^{(\lambda)}$ it follows that the left eigenvector is just the matrix transpose of the right one.
It can be shown that both types of eigenvectors satisfy the bi-orthonormality condition ${\scal{{\kappa_1}_{\rm L}^{(\lambda)}}{{\kappa_2}_{\rm R}^{(\lambda)}}=\delta_{\kappa_1\kappa_2}}$.

We will assume that $\epsilon\I{=}{\rm Re}\lambda$, is set constant, while $\gamma\I{=}-{\rm Im}\lambda$ is varied within the domain ${\gamma\geq0}$.
We can obviously write $\hat{H}^{(\epsilon-i\gamma)}=\hat{H}^{(\epsilon-i0)}-i\gamma\hat{P}_{\rm D}$, where we explicitly introduce a shifted Hermitian Hamiltonian ${\hat{H}^{(\epsilon-i0)}=\hat{H}^{(0)}+\epsilon\hat{P}_{\rm D}}$ and an anti-Hermitian decay-inducing Hamiltonian ${\hat{H}_{\rm D}^{(0-i\gamma)}=-i\gamma\hat{P}_{\rm D}}$.
Evolutions of decay widths for several model settings, averaged over a sample of GOE realizations $N_{\rm R}$, are depicted in Fig.\,\ref{bands}.
We see that the instability expressed by positive decay widths $\Gamma_{\kappa}^{(\lambda)}$ characterizes all eigenstates of $\hat{H}^{(\lambda)}$ as soon as $\gamma$ deviates from zero to infinitesimally small positive values.
Indeed, all eigenstates are expected to have non-zero overlaps with a randomly generated decaying subspace ${\cal H}_{\rm D}$.
Using the Hellmann-Feynman formula, which in the non-Hermitian context reads $\frac{d}{d\lambda}{\cal E}_\kappa^{(\lambda)}\I{=}{\matr{\kappa_{\rm L}^{(\lambda)}}{\frac{d}{d\lambda}\hat{H}^{(\lambda)}}{\kappa_{\rm R}^{(\lambda)}}}$, we derive the following relations: $\frac{\partial}{\partial\gamma}E_{\kappa}^{(\lambda)}\I{=}{-{\rm Im}\matr{\kappa_{\rm L}^{(\lambda)}}{\hat{P}_{\rm D}}{\kappa_{\rm R}^{(\lambda)}}}$ and $\frac{\partial}{\partial\gamma}\Gamma_{\kappa}^{(\lambda)}\I{=}{{\rm Re}\matr{\kappa_{\rm L}^{(\lambda)}}{\hat{P}_{\rm D}}{\kappa_{\rm R}^{(\lambda)}}}$.
These can be used for an estimation of $E_{\kappa}^{(\lambda)}$ and $\Gamma_{\kappa}^{(\lambda)}$ for small values of $\gamma$, where eigenvectors $\ket{\kappa_{\rm R}^{(\lambda)}}$ and $\ket{\kappa_{\rm L}^{(\lambda)}}$ roughly coincide with eigenvectors $\ket{\kappa^{(\epsilon-i0)}}$ of the Hermitian Hamiltonian $\hat{H}^{(\epsilon-i0)}$.
We get 
\begin{eqnarray}
E_{\kappa}^{(\epsilon-i\gamma)}&\approx&E_{\kappa}^{(\epsilon-i0)}+O(\gamma^2)
\label{Esmall}\,,\\
\Gamma_{\kappa}^{(\epsilon-i\gamma)}&\approx&\gamma P_{\kappa}+O(\gamma^2)
\label{Gsmall}\,,
\end{eqnarray}
where ${P_{\kappa}=\sum_{l=1}^{n}|\scal{\phi_l}{\kappa^{(\epsilon-i0)}}|^2}$ is the probability of identifying the eigenstate  $\ket{\kappa^{(\epsilon-i0)}}$ with any of the decaying states $\ket{\phi_l}$.

On the other hand, it is clear that for very large values of ${\gamma>0}$ the decay-inducing part of the Hamiltonian dominates, so that the eigenvectors of the full Hamiltonian \eqref{H} become approximately those of the second term. 
Hence for asymptotic $\gamma$ only ${n<d}$ eigenstates (those roughly coinciding with the set ${\{\ket{\phi_l}\}_{l=1}^{n}\I{\in}{\cal H}_{\rm D}}$) will yield non-negligible widths $\Gamma_{\kappa}^{(\lambda)}\I{\approx}\gamma$, while the rest of ${n_{\bot}=d-n}$ eigenstates (those approximated by $\{\ket{\phi_{\bot l'}}\}_{l'=1}^{n_{\bot}}\I{\in}{\cal H}_{\bot}$) will have ${\Gamma_{\kappa}^{(\lambda)}\I{\approx}0}$.
Indeed, the Hellmann-Feynman formula gives $\frac{\partial}{\partial\gamma}E_{\kappa}^{(\lambda)}\I{\approx}0$ for all eigenstates, and $\frac{\partial}{\partial\gamma}\Gamma_{\kappa}^{(\lambda)}\I{\approx}1$ or 0, depending on whether the approximate eigenvectors belong to ${\cal H}_{\rm D}$ or ${\cal H}_{\bot}$, respectively.
A more systematic treatment makes use of a transformed Hamiltonian $i\gamma^{-1}\hat{H}^{(\epsilon-i\gamma)}=\hat{P}_{\rm D}+i\gamma^{-1}\hat{H}^{(\epsilon-i0)}$ and a perturbative expansion of its complex eigenvalues in small parameter $\gamma^{-1}$.
This yields:
\begin{eqnarray}
E_{\kappa}^{(\epsilon-i\gamma)}&\I{\approx}&E_{\kappa}^{(\epsilon-i\infty)}\I{+}O(\gamma^{-2})
\label{Elarge}\,,\\
\Gamma_{\kappa}^{(\epsilon-i\gamma)}&\I{\approx}&
\left\{\begin{array}{ll}
\gamma\I{+}c_\kappa\gamma^{-1}\I{+}O(\gamma^{-3})& {\rm for\ }\ket{\kappa^{(\epsilon-i\infty)}}\I{\in}{\cal H}_{\rm D},\\
c_\kappa\gamma^{-1}\I{+}O(\gamma^{-3})& {\rm for\ }\ket{\kappa^{(\epsilon-i\infty)}}\I{\in}{\cal H}_{\bot},
\end{array}\right.
\quad\
\label{Glarge}
\end{eqnarray}
where $c_{\kappa}$ are some coefficients.
The log-log dependencies in Fig.\,\ref{bands} support these conclusions.

Having understood the evolution of the spectrum for very small and very large values of $\gamma$, we ask what drives the transition between these limiting regimes.
The region in parameter $\gamma$ where the bifurcation of decay widths takes place is scrutinized under a specific magnifying glass  in Fig.\,\ref{slopes}.
The vertical axis of this figure shows the log-log slope $\chi_{\kappa}^{(\lambda)}\I{=}\frac{d}{d\log\gamma}\log\Gamma_{\kappa}^{(\lambda)}\I{=}(\gamma/\Gamma_{\kappa}^{(\lambda)})\frac{d}{d\gamma}\Gamma_{\kappa}^{(\lambda)}$ of individual decay widths from the $\epsilon\I{=}0$ panel of Fig.\,\ref{bands}.
The value $\chi_{\kappa}^{(\lambda)}\I{=}+1$ indicates the $\Gamma_{\kappa}^{(\lambda)}\I{\propto}\gamma$ behavior, while $\chi_{\kappa}^{(\lambda)}\I{=}-1$ implies $\Gamma_{\kappa}^{(\lambda)}\I{\propto}\gamma^{-1}$.
We observe that the transitional region between these values is a relatively narrow interval of $\gamma$, for the selected values of $d$ and $n$ roughly given by ${1\lesssim\gamma\lesssim 20}$.
In this region, the slopes in Fig.\,\ref{slopes} show large fluctuations, indicating a kind of \uvo{turbulent} evolution of individual level widths.
It turns out that the exceptional points play a crucial role in this evolution.

The EP of a parameter-dependent Hamiltonian $\hat{H}^{(\lambda)}$ is a point $\lambda_{\rm EP}$ of the complex parameter space where two (or more) complex eigenvalues ${\cal E}_{\kappa}^{(\lambda)}$ and ${\cal E}_{\kappa'}^{(\lambda)}$ become degenerate.
In a typical case, the complex degeneracy has a different character than an ordinary degeneracy of a Hermitian Hamiltonian.
First, $\hat{H}^{(\lambda)}$ at ${\lambda=\lambda_{\rm EP}}$ fails to provide a complete system of left and right eigenvectors, but the single eigenvector associated with both degenerate levels becomes self-orthogonal: ${\scal{\kappa_{\rm L}^{(\lambda_{\rm EP})}}{\kappa_{\rm R}^{(\lambda_{\rm EP})}}=0}$.
Second, the EP is not just a conical intersection of the two energy surfaces but a complex square-root singularity.
The path in ${\lambda\in\mathbb{C}}$ satisfying the condition ${{\rm Im}\,{\cal E}_{\kappa}^{(\lambda)}={\rm Im}\,{\cal E}_{\kappa'}^{(\lambda)}}$ is terminated at  $\lambda_{\rm EP}$, where it smoothly connects to herefrom issuing path defined by ${{\rm Re}\,{\cal E}_{\kappa}^{(\lambda)}={\rm Re}\,{\cal E}_{\kappa'}^{(\lambda)}}$.
On the respective sides of the EP, the real and imaginary parts of ${{\cal E}_{\kappa}^{(\lambda)}\I{-}{\cal E}_{\kappa'}^{(\lambda)}}$  bifurcate according to the square-root dependence, which---with regard to the imaginary part---hints at the NHSR behavior.
A passage in a vicinity of an EP induces an \uvo{avoided crossing} of complex eigenvalues (which may include actual crossings of real energies or decay widths) for a single pair of levels along with fast modifications of the corresponding eigenvectors.
The closer is the EP to the selected trajectory of parameter $\lambda$, the sharper are the associated changes.
Multiple EPs induce a turbulent flow of eigenvalues with many avoided crossings and ongoing structural redistributions of eigenvectors, while an absence of EPs near the trajectory implies a laminar flow of eigenvalues and virtual freeze of eigenvectors.

Features of EPs are usually first elucidated on $2\times 2$ matrices.
We sketch such a trivial ${d=2}$ version of our model in Appendix~\ref{ApA} (cf.\,Ref.\,\cite{Hei98}).
It is shown that setting ${\epsilon={\rm Re}\lambda_{\rm EP}}$, so that the path with increasing $\gamma$ crosses the EP of the Hamiltonian at ${\gamma=|{\rm Im}\lambda_{\rm EP}|}$, we observe a sudden bifurcation of the widths $\Gamma_{1}^{(\lambda)}$ and $\Gamma_{2}^{(\lambda)}$ from a common value equal to $\frac{1}{2}\gamma$ to distinct---decreasing and increasing---components.
This is a non-analytic realization of the NHSR in the ${d=2}$ system.
If we increase the distance of $\epsilon$ from ${\rm Re}\lambda_{\rm EP}$, the decay widths of both levels differ already at small $\gamma$ so that the ratio between the widths of short- and long-living state increases with ${|\epsilon\I{-}{\rm Re}\lambda_{\rm EP}|}$.
At the beginning, for $\gamma$ not exceeding $|{\rm Im}\lambda_{\rm EP}|$, both widths grow proportionally to $\gamma$. 
However, for larger $\gamma$, somewhere at $\gamma\gtrsim|{\rm Im}\lambda_{\rm EP}|$, the smaller of both widths turns back and starts decreasing as $\gamma^{-1}$. 
This is a smooth but still clear realization of the NHSR effect.
Examples of these scenarios, depicted in Fig.\,\ref{2D} below, can be compared to less trivial but rather analogous behavior for ${d>2}$ shown in Fig.\,\ref{bands}.

\begin{figure}[t!]
\includegraphics[width=\linewidth]{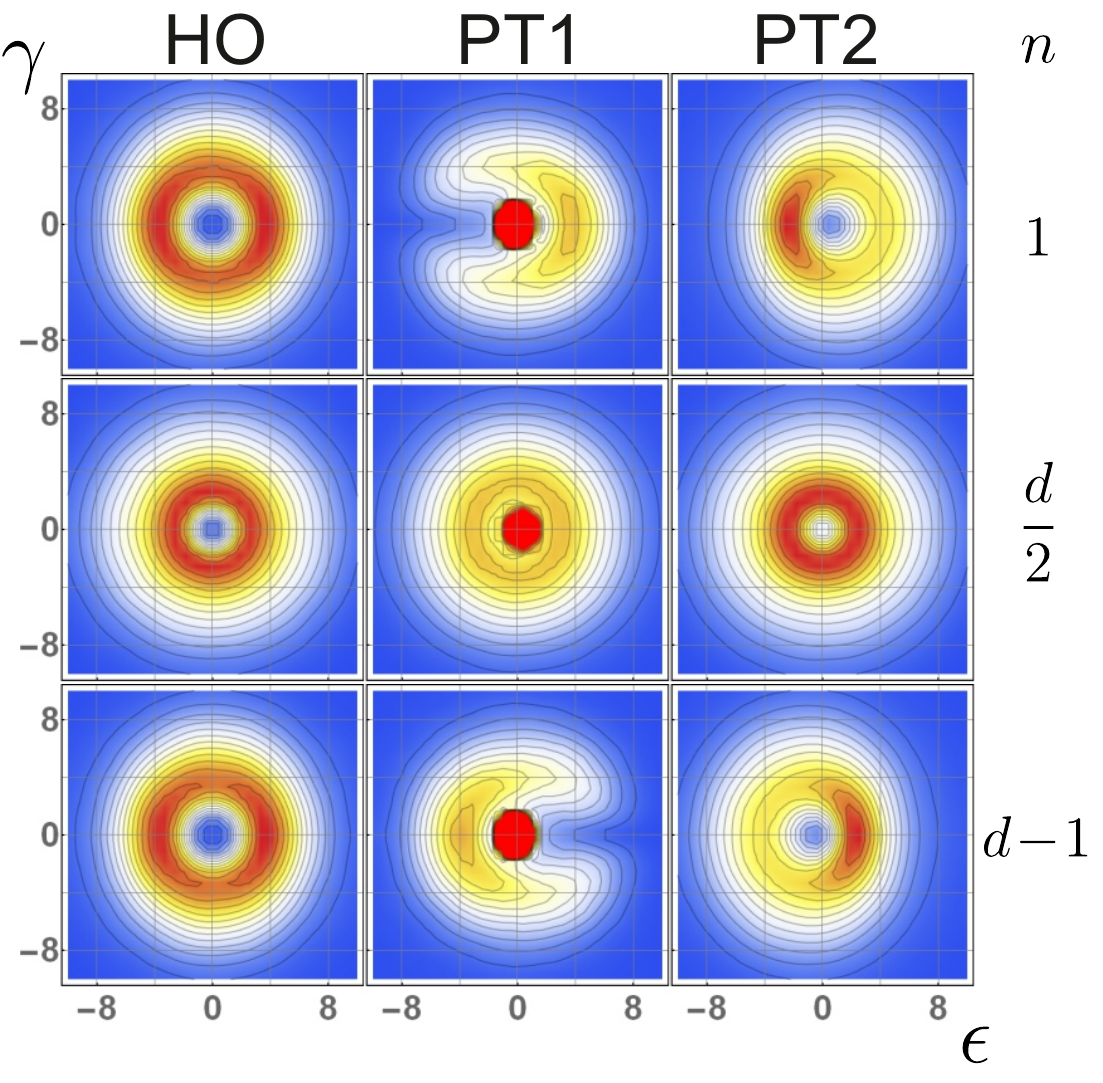}
\caption{
Distributions of EPs in the complex plane of parameter $\lambda$ for Hamiltonians \eqref{H} with the unperturbed term $\hat{H}^{(0)}$ equal to the forms \eqref{H0ho}, \eqref{H0pt1} and \eqref{H0pt2} (the columns \uvo{HO}, \uvo{PT1} and \uvo{PT2}, respectively) and the decaying subspace dimensions set to ${n=1}$, ${n=\frac{1}{2}d}$ and ${n=d-1}$ (the rows).
In all cases ${d=16}$.
Warmer colors indicate larger EP densities.
Each panel comprises a total number $N_{\rm EP}N_{\rm R}\I{\approx}10^6$ of generated EPs [the respective $N_{\rm R}$ follows from Eq.\,\eqref{Nep}].}
\label{distri}
\end{figure}   

Focusing on these ${d>2}$ cases, we first note that a symmetric matrix of dimension $d$ depending linearly on parameter ${\lambda\in{\mathbb C}}$ has in general $\frac{1}{2}d(d\I{-}1)$ pairs of complex-conjugate EPs.
However, in case of Hamiltonian \eqref{H} a large part of EP pairs  migrates to infinity, while at finite values of $\lambda$ we observe a reduced number of pairs:
\begin{equation}
\tfrac{1}{2}N_{\rm EP}=n(d\I{-}n)
\,.\label{Nep}
\end{equation} 
This is $\lesssim 50\,\%$ of the full number of pairs (the maximal fraction is reached for ${n=\frac{1}{2}d}$).
Figure~\ref{distri} shows distributions of these EPs in ${\lambda\in{\mathbb C}}$ for the three initial Hamiltonians \eqref{H0ho}--\eqref{H0pt2} and for various sizes of the decaying subspace ${\cal H}_{\rm D}$.
The distribution is averaged over a large number of GOE realizations of ${\cal H}_{\rm D}$.

The Hilbert space dimension in Fig.\,\ref{distri}  is rather moderate, $d=16$, but the distributions of EPs remain qualitatively similar for larger dimensions, except an overall scaling of the absolute value $|\lambda|$.  
The essential fraction of the EP distribution is expected to be located within a domain defined by
\begin{equation}
|\lambda|\I{\lesssim}\frac{S\,d^2}{\sqrt{n(d\I{-}n)}}\I{\propto}
\left\{\begin{array}{ll}
d^{3/2} & {\rm for\ } n\I{\ll}d {\ \rm or\ } n_{\bot}\I{\ll}d, \\
d  & {\rm for\ } n\I{\approx}n_{\bot}\I{\approx}d/2,
\end{array}\right.
\label{scala}
\end{equation}
where $S\I{=}\bigl(\overline{\Delta^2 E}^{(0)}/d^2\bigr)^{1/2}$ is the linear spread of the unperturbed spectrum divided by the dimension (e.g., for $H^{(0)}_{\rm HO}$ and $H^{(0)}_{\rm PT2}$ with unit average spacings we get $S\I{\approx}0.3$).
The scaling formula \eqref{scala} follows from the analysis of Ref.\,\cite{Str18} (cf.\,\cite{Zir83}) which showed that a maximum of the EP distribution for a general perturbed Hamiltonian ${{\hat H}^{(0)}+\lambda\hat{V}}$ is achieved when spectral variances (spreads of the spectra) of both terms $H^{(0)}$ and $\lambda\hat{V}$ are approximately equal to each other.
The formula is also confirmed by our numerical simulations.

\begin{figure}[t!]
\includegraphics[width=\linewidth]{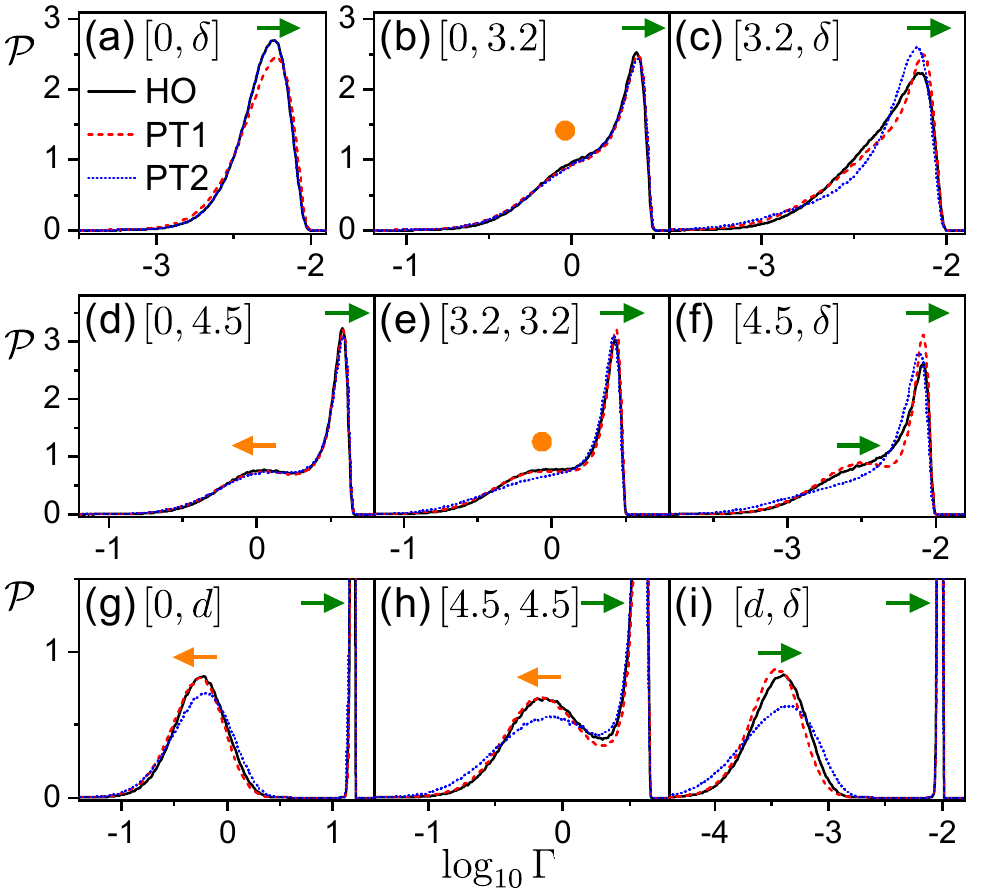}
\caption{
Distribution of decay widths for the model \eqref{H} with ${d=2n=16}$ at various values of ${\lambda=\epsilon-i\gamma}$ indicated in each panel in parentheses $[\epsilon,\gamma]$ (some panels correspond to vertical lines in Figs.\,\ref{bands} and \ref{slopes}).
Averaging over $N_{\rm R}\I{\approx}10^{6}/d$ of GOE realizations is performed. 
The density ${\cal P}$ is normalized so that $\int{\cal P}\,d(\log_{10}\!\Gamma)\I{=}1$.
Panel (a) corresponds to ${\lambda\approx 0}$ (${\delta=0.01}$), panels (b) and (c) are associated with ${|\lambda|\approx 3.2}$, panels (d), (e) and (f) with ${|\lambda|\approx 4.5}$, and panels (g), (h) and (i) represent ${|\lambda|>6}$ cases.
The arrows indicate the direction of motions of the respective parts of the distribution with increasing $\gamma$ (dots denote parts that are just at the start of motion).
The three choices of initial Hamiltonian are distinguished by different line types.
For the corresponding distributions of EPs see the medium row of Fig.\,\ref{distri}.
}
\label{widths1}
\end{figure}   

Let us proceed to the discussion of the impact of the EP distribution on the NHSR dynamics.
We saw in Fig.\,\ref{bands} (the lower right panel) that two distinct groups of states with shorter and longer lifetimes can exist already long before the superradiant transition takes place, i.e., below the $\gamma$ value where the longer-living group turns back to the $\Gamma\I{\to}0$ path.
So in the investigation of the NHSR we have to distinguish a {\em static\/} bi-modality (existence of two peaks in the distribution of decay widths which both move with increasing $\gamma$ to the increasing-$\Gamma$ direction) from a {\em dynamic\/} bi-modality (existence of two peaks moving to opposite $\Gamma\I{\to}0$ and $\Gamma\I{\to}\infty$ directions).
The EPs play an important---though not fully deterministic---role in the description of both these features.
The static bimodality is achieved if the value of $\epsilon$ either exceeds the upper peripheral region of the EP distribution projected to the ${\rm Re}\lambda$ axis, or undershoots its lower peripheral region.
Analogously, the dynamic bimodality occurs when the value of $\gamma$ passes the upper periphery of the EP distribution projected of  to the ${\rm Im}\lambda$ axis.
However, it needs to be stressed that details of the formation of both kinds of bi-modalities for $\lambda$ located still within the principal part of the EP distribution are certainly beyond the above simplified description.

\begin{figure}[t!]
\includegraphics[width=\linewidth]{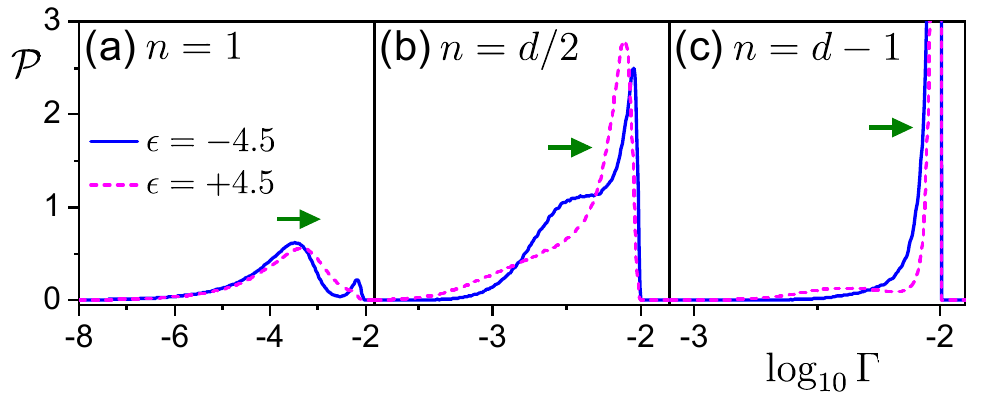} 
\caption{
Distribution of decay widths for ${H^{(0)}=H^{(0)}_{\rm PT2}}$ at ${\epsilon=\pm 4.5}$ (the full and dashed curves) and ${\gamma=\delta=0.01}$. 
The calculation is done for ${d=16}$ with $N_{\rm R}d\I{\approx}10^{6}$.  
The three panels correspond to (a) ${n=1}$, (b) ${n=8}$, and (c) ${n=15}$; cf. the EP distributions in the last column of Fig.\,\ref{distri}.
}
\label{widths+-}
\end{figure}   

To support the claims from the previous paragraph, we show in Fig.\,\ref{widths1} several transient distributions of decay widths for the same model setup as used in the middle row of Fig.\,\ref{distri} (${d=2n=16}$).
Each panel depicts a distribution of widths averaged over a large number of GOE realizations for a fixed parameter point ${\lambda=\epsilon-i\gamma}$ together with arrows, which indicate the direction in which a given part of the distribution evolves with increasing $\gamma$. 
Note that some of the distributions represent $\gamma$=const.\,cuts of Figs.\,\ref{bands} and \ref{slopes}.
The fixed parameter points in Fig.\,\ref{widths1} are organized so that panel (a) is associated with $\lambda$ close to zero, while the pair of panels (b),(c) and further the triples (d),(e),(f) and (g),(h),(i) correspond to increasing distance from the origin in various directions of the ${\lambda\in{\mathbb C}}$ plane.
We see that the panels within the same group show a comparable degree of splitting of the spectrum to short- and long-living states, but the bi-modality has the static character if $\gamma$ is not large enough.
The dynamic bi-modality is triggered only when $\gamma$ passes the upper periphery of the projected EP distribution.
 
The distribution of EPs in Fig.\,\ref{distri} for ${n=\frac{1}{2}d}$, which is the case corresponding to Fig.\,\ref{widths1}, is approximately symmetric under rotations around ${\lambda=0}$.
However, the distributions for $n\I{=}1$ and $n=d\I{-}1$ with ${H^{(0)}=H^{(0)}_{\rm PT1}}$ and $H^{(0)}_{\rm PT2}$ exhibit strong asymmetries under rotations and even under the mirror transformation $\epsilon\leftrightarrow-\epsilon$.
In these cases, the NHSR dynamics sensitively depends on the sign of $\epsilon$. 
This is illustrated in Fig.\,\ref{widths+-}, where we show the decay width distributions at two points with ${\rm Re}\lambda\I{=}\pm\epsilon$ and ${\rm Im}\lambda$ small for $H^{(0)}_{\rm PT2}$. 
Both points belong to the regions within the principal parts of the EP distributions in the rightmost column of Fig.\,\ref{distri}.
We see that in the $n\I{=}1$ and $n=d\I{-}1$ cases the decay width distributions have significantly different forms for both $\pm\epsilon$ values, in agreement with the observed asymmetries of the respective EP distributions.
In both these cases, the width distribution corresponding to the $\epsilon$ value closer to the main maximum of the EP distribution shows a more developed static bi-modality.
However, the  width distributions for $\pm\epsilon$ differ also in the $n\I{=}\frac{1}{2}d$ case, for which the EP distribution is roughly symmetric.
This indicates that initial stages of the static bi-modality formation transcend the description based on the overall EP distribution, which does non reflect links of individual EPs to specific pairs of levels (cf.\,Ref.\,\cite{Str18}).

\section{Global properties of the complex spectrum}
\label{GLOPR}

Employing the trace-based method described e.g. in Ref.\,\cite{Str18} (see also \cite{Wol80}), one can derive expressions for cumulants of the whole eigenvalue spectrum of a general (parameter-dependent) Hamiltonian $\hat{H}^{(\lambda)}$.
In particular, the average of all eigenvalues is determined from
\begin{equation}
\ove{{\cal E}}^{(\lambda)}\equiv\frac{1}{d}\sum_{\kappa=1}^d{\cal E}_{\kappa}^{(\lambda)}=\frac{1}{d}\,{\rm Tr}\hat{H}^{(\lambda)}
\label{defave}
\,,
\end{equation}
and the eigenvalue variance is given by 
\begin{eqnarray}
\ove{\Delta^2{\cal E}}^{(\lambda)}&\equiv&\frac{1}{d}\sum_{\kappa=1}^d\bigl[{\cal E}_{\kappa}^{(\lambda)}\I{-}\ove{{\cal E}}^{(\lambda)}\bigr]^2=
\nonumber\\
&=&\frac{1}{d}\,{\rm Tr}{\hat{H}^{(\lambda)}}{^2}-\frac{1}{d^2}{\rm Tr}^2\hat{H}^{(\lambda)}
\label{defdis}
\,.
\end{eqnarray}
For the non-Hermitian Hamiltonian \eqref{H}, the averages and variances related to complex energies ${\cal E}_{\kappa}^{(\lambda)}$ shall be evaluated separately for real and imaginary parts.
From Eq.\,\eqref{defave} we get easy formulas for both types of averages,
\begin{eqnarray}
\ove{E}^{(\epsilon-i\gamma)}-\ove{E}^{(0)}&=&\frac{n}{d}\epsilon
\label{aveE}
\,,
\\
\ove{\Gamma}^{(\epsilon-i\gamma)}&=&\frac{n}{d}\gamma
\label{aveG}
\,,
\end{eqnarray}
and from Eq.\,\eqref{defdis} we obtain the following slightly more complicated relations for the variances:
\begin{equation}
\ove{\Delta^2E}^{(\epsilon-i0)}\!\!-\ove{\Delta^2E}^{(0)}=2\epsilon\,\frac{n}{d}\,A+\epsilon^2\,\frac{n}{d}\left(1\I{-}\frac{n}{d}\right)
\label{disre},\quad
\end{equation}
where ${A=\frac{1}{n}\sum_{k=1}^{d}\bigl(\!E_{k}^{(0)}\!\I{-}\ove{E}^{(0)}\bigr)\matr{k}{\hat{P}_{\rm D}}{k}}$, and
\begin{equation}
\ove{\Delta^2E}^{(\epsilon-i\gamma)}\!\!-\ove{\Delta^2E}^{(\epsilon-i0)}=\ove{\Delta^2\Gamma}^{(\epsilon-i\gamma)}\!\!-\gamma^2\,\frac{n}{d}\left(1\I{-}\frac{n}{d}\right)
\label{disim}.
\end{equation}
Formulas \eqref{aveE} and \eqref{aveG} imply linear dependences of the average real energy and the average decay width on $\epsilon$ and $\gamma$, respectively. 
Formula \eqref{disre} expresses a quadratic dependence of the real energy variance on $\epsilon$ for ${\gamma=0}$, while formula \eqref{disim} captures a specific relation between the variances of real energies and decay widths for variable $\gamma$ and constant $\epsilon$.

The linear term of the quadratic dependence of the real energy variance in Eq.\,\eqref{disre} is proportional to the coefficient $A$, which quantifies an asymmetry of the unperturbed energy spectrum with respect to the decaying subspace. 
It is evaluated as an average of the energy displacement ${(\!E_{k}^{(0)}\!\I{-}\ove{E}^{(0)})}$ calculated with normalized weight factors ${w_k=\frac{1}{n}\sum_{l=1}^{n}|\scal{\phi_l}{k}|^2}$ proportional to the overlap probability of the respective unperturbed eigenstate with ${\cal H}_{\rm D}$.
This asymmetry determines the point ${\epsilon_{\rm min}=-Ad/(d\I{-}n)}$ of a minimal quadratic spread $\ove{\Delta^2E}^{(\epsilon-i0)}$ of the real spectrum along the real $\lambda$ axis, i.e., the point of maximal compression of the spectrum.
This point shall roughly correspond to the projection of the \uvo{center of mass} of the EP distribution to the real $\lambda$ axis.
The coefficient $A$ depends on the statistical realization of the decaying subspace, as well as on the energy spectrum of the initial Hamiltonian.
The GOE average is obviously $\ave{A}\I{=}0$, but an analytic evaluation of the higher cumulants is hindered by non-trivial correlations of weight factors $w_k$ for various $k$.
Nevertheless, we can infer that $\ave{\Delta^2A}$ and $\ave{\Delta^3A}$ are correlated with the respective cumulants of the unperturbed energy spectrum $\ove{\Delta^2E}^{(0)}$ and $\ove{\Delta^3E}^{(0)}$.
This qualitatively explains the above-mentioned $\epsilon\leftrightarrow-\epsilon$ asymmetries of EP distributions in Fig.\,\ref{distri} for ${H^{(0)}\I{=}H^{(0)}_{\rm PT1}}$ and $H^{(0)}_{\rm PT2}$, whose spectra are apparently skewed (cf.\,Fig.\,\ref{spek}).

\begin{figure}[t!]
\includegraphics[width=\linewidth]{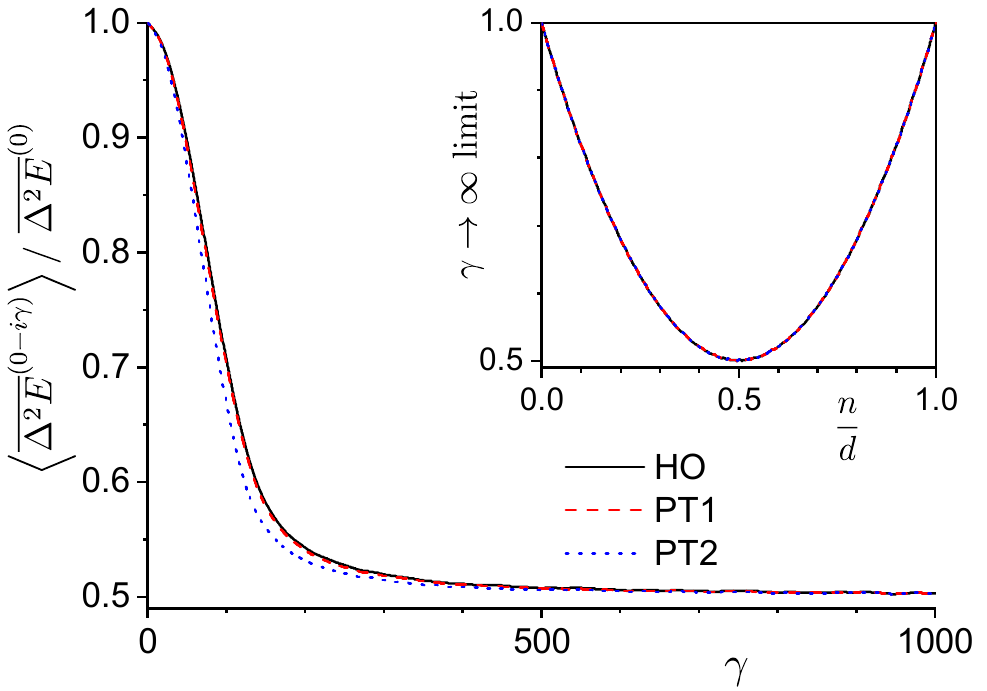} 
\caption{
Decrease of the squared spread (variance) of the real energy spectrum $\overline{\Delta^2 E}^{(\lambda)}$ with increasing $\gamma$ for $\epsilon\I{=}0$. 
The picture was obtained by averaging over $N_{\rm R}\I{=}64$ realizations for $d\I{=}2n\I{=}256$. 
The variance is normalized to the value at $\lambda\I{=}0$.
The inset shows the asymptotic value of the relative variance as a function of $n/d$, which closely follows an empirical formula $\overline{\Delta^2 E}^{(0-i\infty)}/\overline{\Delta^2 E}^{(0)}\approx 1\I{-}2\frac{n}{d}(1\I{-}\frac{n}{d})$.
}
\label{contra}
\end{figure}   

We now turn to the consequences of Eq.\,\eqref{disim} for the NHSR dynamics.
The left-hand side of this formula represents the change of variance of the real spectrum between $\gamma\I{=}0$ and $\gamma\I{>}0$ points with the same $\epsilon$.
The right-hand side is the difference between the actual variance of the decay widths at ${\lambda=\epsilon-i\gamma}$ and the variance of a two-component ensemble consisting of $n$ widths equal to $\gamma$ and ${n_{\bot}=d-n}$ widths equal to 0. 
According to Ref.\,\cite{Wol80}, the spectrum of Hamiltonian \eqref{H} satisfies the following inequalities:
\begin{eqnarray}
E_{1}^{(\epsilon-i0)}\leq{\rm Re}\,{\cal E}_{\kappa}^{(\epsilon-i\gamma)}&=&E_{\kappa}^{(\epsilon-i\gamma)}\leq E_{d}^{(\epsilon-i0)}
\,\label{ineq1}\\
0\leq-{\rm Im}\,{\cal E}_{\kappa}^{(\epsilon-i\gamma)}&=&\Gamma_{\kappa}^{(\epsilon-i\gamma)}\leq\gamma
\,.\label{ineq2}
\end{eqnarray}
The bounds $E_{1}^{(\epsilon-i0)}$ and $E_{d}^{(\epsilon-i0)}$ in \eqref{ineq1} are the lowest and the highest eigenvalues of $\hat{H}^{(\epsilon-i0)}$, and similarly 0 and $\gamma$ in \eqref{ineq2} are the lowest and highest eigenvalues of  ${\gamma\hat{P}_{\rm D}}$.
Inequality \eqref{ineq2} implies that the right-hand side of Eq.\,\eqref{disim} is semi-negative.
The limiting zero value corresponds to ${\gamma=0}$, while for increasing $\gamma$ we expect negative values that remain of order ${\sim O(1)}$ even in the asymptotic case. 
Hence we see that the increase of the decay rate $\gamma$ always reduces the spread of the real energy spectrum.
This reduction is correlated with the increase of the spread of decay widths of individual eigenstates and survives the ${\gamma\to\infty}$ limit.
As shown in Fig.\,\ref{contra}, the maximal contraction of the real energy spectrum, reaching the asymptotic value of 50\,\% of the squared spread of the original spectrum (${\approx 71\,\%}$ of the linear spread), is obtained for ${n=\frac{1}{2}d}$.

\section{Effects of criticality}
\label{CRIH}

\begin{figure*}[t!]
\includegraphics[width=0.8\linewidth]{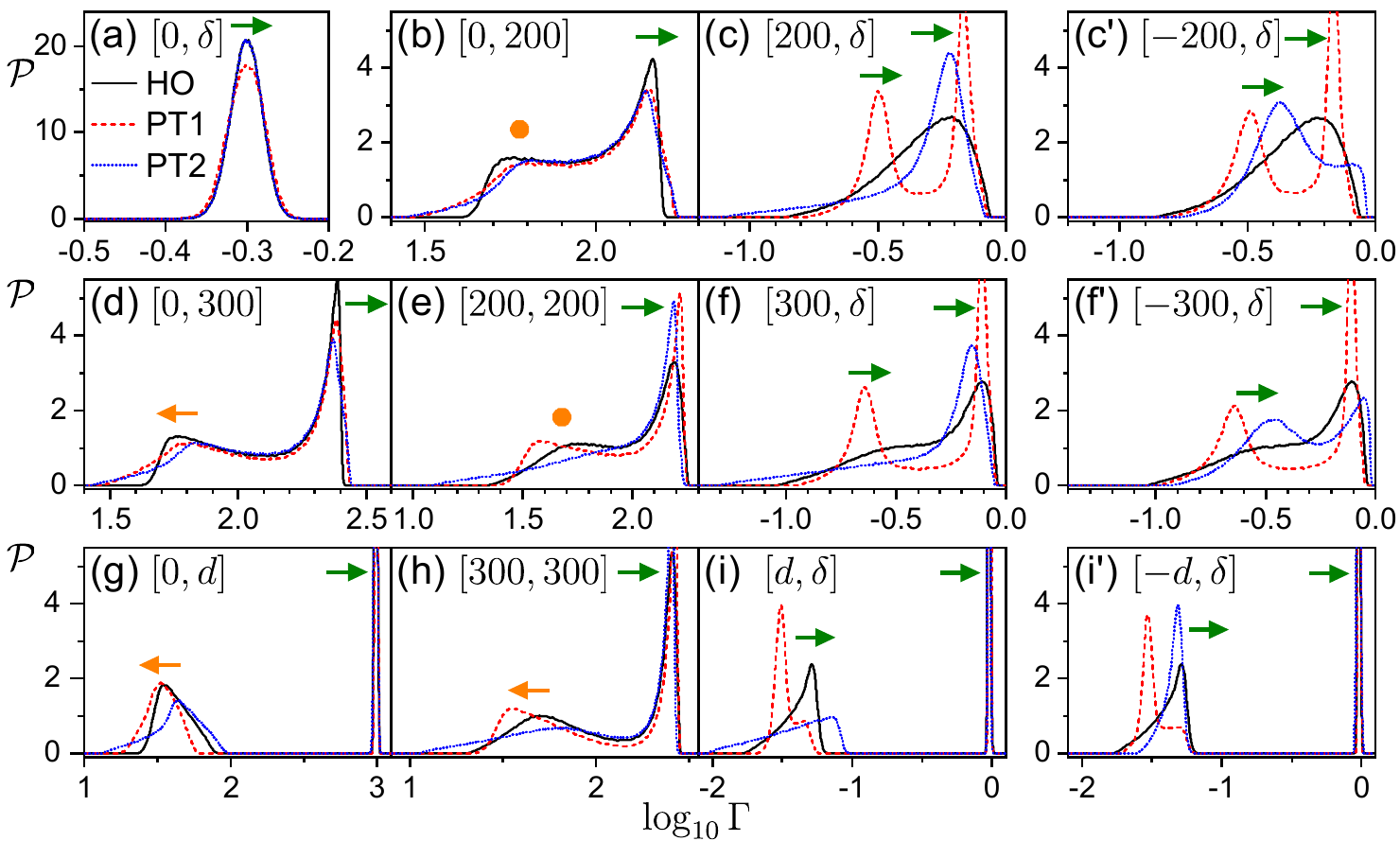} 
\caption{
Distribution of decay widths at various points $[\epsilon,\gamma]$ for ${d=2n=1024}$.
The number of realizations $N_{\rm R}\I{\approx}10^3$.
Compared to Fig.\,\ref{widths1}, the parameter values $[\epsilon,\gamma]$ in panels (a)--(i) are enlarged by a factor roughly equal to the fraction of dimensions 1024/16, cf. formula \eqref{scala}.
Additional panels (c'), (f') and (i') show width distributions for negative values of $\epsilon$ at mirror-imagined places with respect to panels (c), (f), and (i), respectively.
}
\label{widths2}
\end{figure*}   

\begin{figure}[t!]
\includegraphics[width=\linewidth]{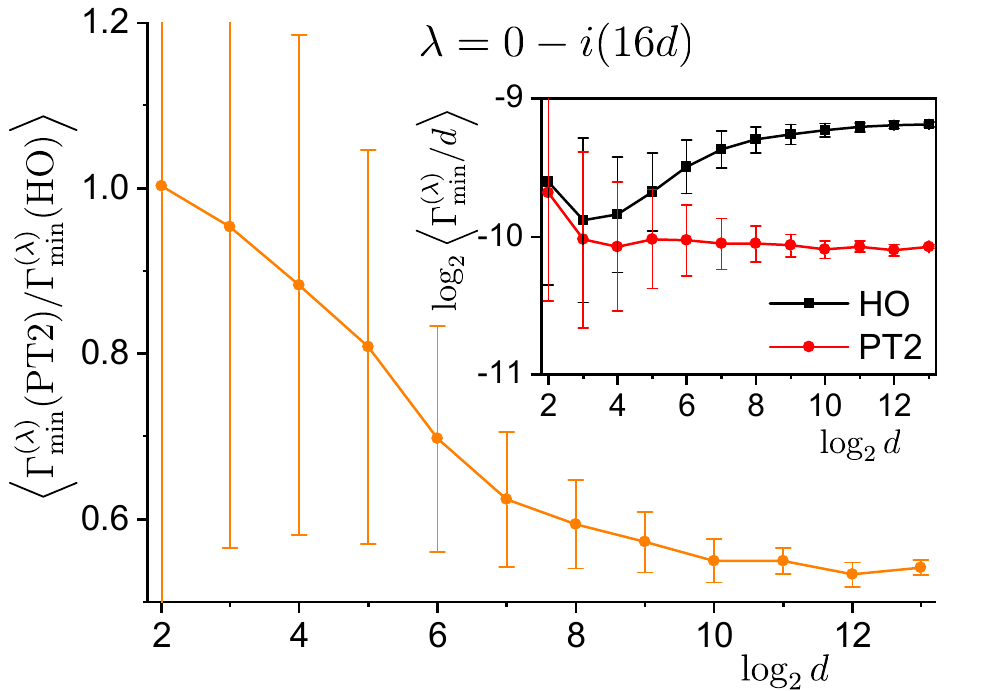} 
\caption{
Comparison of the minimal decay widths in the whole set of levels for $\hat{H}^{(0)}\I{=}\hat{H}^{(0)}_{\rm HO}$ and $\hat{H}^{(0)}_{\rm PT2}$ at $[\epsilon,\gamma]\I{=}[0,16d]$.
The dimension $d$ is varied and $n\I{=}\frac{1}{2}d$.
The inset shows both minimal widths separately, the main panel depicts their ratio.
The dots represent GOE averages obtained in $N_{\rm R}\I{\approx}6.5\I{\cdot}10^4/d$ realizations, error bars indicate $1\sigma$ standard deviations.  
}
\label{lines}
\end{figure}

As seen in Fig.\,\ref{widths1}, the distributions of decay widths for various choices \eqref{H0ho}, \eqref{H0pt1} and \eqref{H0pt2} of the initial Hamiltonian $\hat{H}^{(0)}$ are rather close to each other for moderate dimensions $d$.
However, the differences get much more significant as $d$ grows.
In Fig.\,\ref{widths2} we show an analog of  Fig.\,\ref{widths1} for ${d=1024}$.
In comparison to the previous figure, the parameter coordinates ${[\epsilon,\gamma]}$ corresponding to panels (a)--(i) were scaled roughly according to Eq.\,\eqref{scala}.
We immediately see that the choice of $\hat{H}^{(0)}$ plays a substantial role.
In general, the first- and second-order QPT Hamiltonians $\hat{H}^{(0)}_{\rm PT1}$ and $\hat{H}^{(0)}_{\rm PT2}$ result at various fixed values of $\lambda$ in much more developed bi-modal structures of the width distribution than the harmonic oscillator Hamiltonian $\hat{H}^{(0)}_{\rm HO}$.
Interestingly, comparing pairs of panels (c)-(c'), (f)-(f') and (i)-(i') in Fig.\,\ref{widths2}, we observe clear differences between decay width distributions for both critical Hamiltonians at mirror-conjugate values of ${\rm Re}\lambda$.
These again reflect the $-\epsilon\I{\leftrightarrow}+\epsilon$ asymmetries discussed in connection with Fig.\,\ref{widths+-}(b) and Eq.\,\eqref{disre}.


The differences of decay width distributions in Fig.\,\ref{widths2} may indicate that for the critical Hamiltonians the NHSR dynamics is more advanced than for the non-critical one.
However, as shown above, the NHSR is not defined by a static separation of short- and long-living states at a given $\gamma$, but more substantially relies on the reversed evolution of both modes as $\gamma$ grows to large values and on the convergence of the long-living mode to $\Gamma\I{=}0$.
Whether the criticality of $\hat{H}^{(0)}$ implies some enhancement of the latter process is studied in Fig.\,\ref{lines}.

This figure compares the smallest decay widths $\Gamma_{\rm min}^{(\lambda)}(\rm HO)$ and $\Gamma_{\rm min}^{(\lambda)}(\rm PT2)$ obtained by minimization over the entire set of all $d$ levels for the harmonic oscillator Hamiltonian $\hat{H}^{(0)}_{\rm HO}$ and the second-order QPT Hamiltonian $\hat{H}^{(0)}_{\rm PT2}$.
The imaginary part of $\lambda$ is taken very large, $\gamma\I{=}16d$, so that we are far above the superradiant transition to decreasing width regime, the real part of $\lambda$ is set to $\epsilon\I{=}0$. 
Results are displayed for several Hilbert space dimensions ranging from $d\I{=}4$ to $d\I{=}8192$, the dimension of the decaying subspace being always $n\I{=}\frac{1}{2}d$.
The minimal decay widths and their ratios were averaged over several GOE realizations, yielding the dots in Fig.\,\ref{lines}.
The error bars indicate standard deviations of the distributions obtained.
Note that the numbers of realizations $N_{\rm R}$ decrease with $d$ for computability reasons, but they are sufficient for the purposes of the present study.
The numerical precision, however, prevents us to do the same calculation for $\hat{H}^{(0)}_{\rm PT1}$, for which the parity doublets of levels become too close with increasing $d$. 

We see in the inset of Fig.\,\ref{lines} that for $d\I{\gtrsim}2^{10}$ both GOE averages of the minimal decay width scale roughly as $\Gamma_{\rm min}^{(\lambda)}\I{\propto}d$.
This is consistent with the choice $\gamma\I{=}16d$ since, with respect to the scaling formula \eqref{scala}, the pattern in  Fig.\,\ref{bands} remains roughly invariant under the change of dimension if variables $\gamma$ and $\Gamma$ are replaced by $\gamma/d$ and $\Gamma/d$, respectively.

The most important information follows from the ratio between both widths for large $d$.
A plot of the GOE average $\ave{\Gamma_{\rm min}^{(\lambda)}({\rm PT2})/\Gamma_{\rm min}^{(\lambda)}({\rm HO})}$ at $\lambda\I{=}0\I{-}i(16d)$ is shown in the main panel of Fig.\,\ref{lines}.
It suggests that the asymptotic value of the average ratio is roughly equal to 0.55.
Since in the large-$\gamma$ domain the evolution of widths is described by Eq.\,\eqref{Glarge}, hence ${\Gamma_{\rm min}^{(\lambda)}\approx c_{\rm min}\gamma^{-1}}$, the above numerical value is assigned also to the average ratio $\ave{c_{\rm min}({\rm PT2})/c_{\rm min}({\rm HO})}$ of the respective coefficients.
It is clear that the particular value 0.55 would not apply to other choices of $\epsilon$ (cf.\,Fig.\,\ref{widths2}), but the NHSR process is generally more advanced for the second-order critical system than for the non-critical harmonic oscillator.

Let us stress that all initial Hamiltonians in this study have the same spread of spectra (average spacing of levels), so the discussed effect reflects some tinier differences in the distributions of levels.
Indeed, Hamiltonian $\hat{H}^{(0)}_{\rm PT2}$ with the quartic potential exhibits a strong (increasing with $d$) accumulation of levels near the lowest energy, so a higher occurrence of small spacings between levels.
Therefore, some EPs appear closer to $\lambda\I{=}0$ and some of the levels exhibit the crossover to the ${\Gamma_{\kappa}^{(\lambda)}\propto\gamma^{-1}}$ regime sooner than in the case of the harmonic oscillator. 
This explains the observed acceleration of the NHSR process for the second-order QPT Hamiltonian. 
There is no doubt that a similar (even stronger) effect would be observed for the first-order critical Hamiltonian $\hat{H}^{(0)}_{\rm PT1}$, if it is made available for large-$d$ numerical simulations.

\section{Summary}
\label{SUMA}

We summarize our main results in the following items:

(i) The non-Hermitian superradiance (i.e., splitting of energy eigenstates of an open quantum system with increasing coupling of a certain subset of states to continuum into the groups of short- and long-living states with decay widths $\Gamma\I{\to}\infty$ and $\Gamma\I{\to}0$) is a universal effect, qualitatively independent of the form of the initial Hamiltonian and the choice of the decaying subspace.
The effect is in a rudimentary form present already in a two-level system (Fig.\,\ref{2D}), which contains the essence of more complicated dependencies for larger dimensions. 
 
(ii) The initial (small-coupling) and asymptotic (large-coupling) stages of the NHSR evolution are understood from elementary perturbative expressions. 
In log-log plots (like Fig.\,\ref{bands}) we observe laminar flows of real energies and decay widths described by the respective pair of equations \eqref{Esmall}--\eqref{Gsmall} or \eqref{Elarge}--\eqref{Glarge}.

(iii) The intermediate stages of the NHSR evolution are strongly influenced by the distribution of EPs of an effective non-Hermitian Hamiltonian.
At these stages, the real energies and decay widths evolve in a turbulent way, showing numerous collisions (actual or avoided crossings) and large fluctuations of log-log slopes (see Fig.\,\ref{slopes}).

(iv) Essential information follows from projections of the EP distribution (cf.\,Fig.\,\ref{distri}) to real and imaginary axes of the coupling parameter $\lambda\I{=}\epsilon\I{-}i\gamma$.
Considering trajectories in the $\lambda$ plane with fixed $\epsilon$, we distinguish two basic ESQPT scenarios (see Fig.\,\ref{bands}):
First, if $\epsilon$ is within the support of the projected EP distribution on the real axis, the decay widths of individual levels for small $\gamma$ form a single-mode distribution. 
As $\gamma$ increases, the distribution moves to increasing values of $\Gamma$ and eventually splits into the increasing and decreasing branches when $\gamma$ gets outside the support of the projected EP distribution on the imaginary axis.
Second, if $\epsilon$ is outside the support of the real projection of the EP distribution, the decay-width distribution is bi-modal already at small values of $\gamma$.
However, the splitting into the increasing and decreasing branches again happens only when $\gamma$ gets out of the imaginary projection of the EP distribution.
These conclusions are supported by examples of $\Gamma$ distributions in Figs.\,\ref{widths1}, \ref{widths+-} and \ref{widths2}, but must be taken only as raw simplifications.  

(v) Squared spreads of the spectra of real energies and decay widths are mutually correlated [see Eq.\,\eqref{disim}].
The opening of the system always leads to narrowing of the real energy spectrum (see Fig.\,\ref{contra}).

(vi) Details of the NHSR dynamics depend on the structure of real energy spectrum of the $\gamma\I{=}0$ Hamiltonian.
First, the formation of bi-modal structures in the decay width distribution depends on the value of $\epsilon$ in a way that goes beyond the overall EP distribution (see examples in  Fig.\,\ref{widths+-}). 
Second, even the asymptotic-$\gamma$ behavior carries traces of the $\gamma\I{=}0$ energy spectrum.
In particular, the presence of small spacings between energy levels is equivalent to a closer approach of some of the EPs to the real $\lambda$ axis, which shifts the crossover of some states to the $\Gamma\I{\to}0$ stage of evolution to smaller values of parameter $\gamma$ and makes the stabilization process of these states more advanced (see Fig.\,\ref{lines}).
In this sense one can observe a speedup of the NHSR effect.
This applies especially to systems at quantum critical points, which contain closely located levels by definition.

\section*{Acknowledgments}
\label{sec:Ack}
We acknowledge funding by the Charles University Project UNCE/SCI/013.

\appendix
\section{Two-dimensional model}
\label{ApA}

\begin{figure}[t!]
\includegraphics[width=\linewidth]{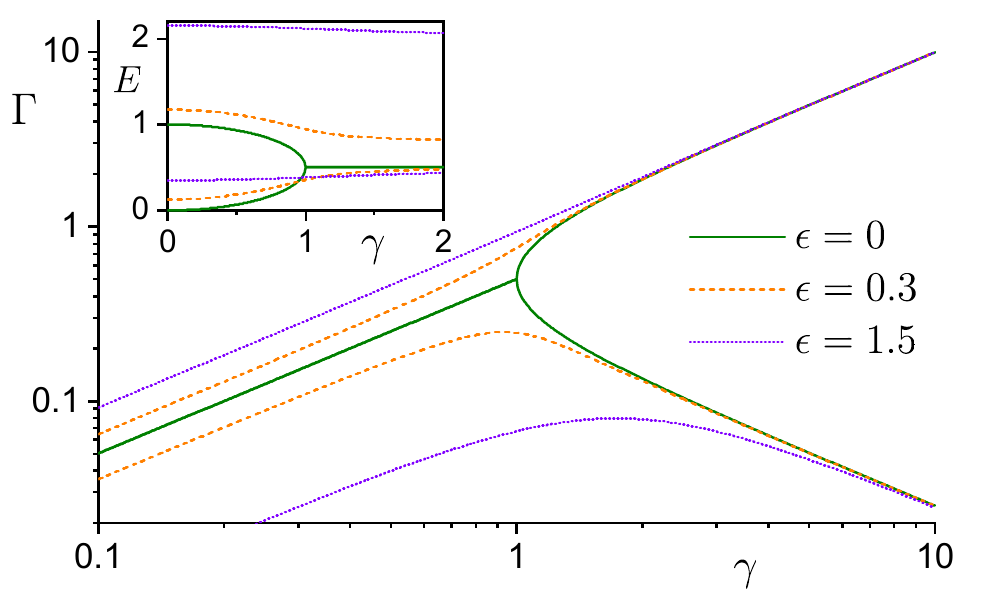}
\caption{
Bifurcations of decay widths for ${d=2}$ model \eqref{H2} with ${E^{(0)}_2\I{-}E^{(0)}_1=1}$ and ${\vartheta=\pi/4}$ for various values of $\epsilon$.
The inset shows the corresponding evolution of real energies.
The case ${\epsilon=0}$ corresponds to the direct passage through the EP, while the pairs of curves with ${\epsilon=0.3}$ and 1.5 represent the effect of the EP on increasingly distant ${\lambda\in{\mathbb C}}$ trajectories.
}
\label{2D}
\end{figure}

Hamiltonian \eqref{H} in the ${d=2}$ case reads
\begin{equation}
\hat{H}^{(\lambda)}=\underbrace{\left(\begin{matrix}E^{(0)}_{1}&0\\0&E^{(0)}_{2}\end{matrix}\right)}_{\hat{H}^{(0)}}+\underbrace{(\epsilon\I{-}i\gamma)}_{\lambda}\underbrace{\left(\begin{matrix}\cos^2\vartheta&\cos\vartheta\sin\vartheta\\\sin\vartheta\cos\vartheta&\sin^2\vartheta\end{matrix}\right)}_{\left(\smallmatrix\cos\vartheta\\\sin\vartheta\endsmallmatrix\right)(\cos\vartheta\ \sin\vartheta)}
\,,
\label{H2}
\end{equation}
where $\vartheta$ (assumingly ${\neq 0,\pi}$ since these special values would yield trivial solutions) is an angle determining the single decaying state $\ket{\phi_1}$. 
Note that for ${d=2}$ only the case of ${n=1}$ is non-trivial.
Two complex eigenvalues
\begin{eqnarray}
{\cal E}_{1,2}^{(\lambda)}\equiv E_{1,2}^{(\lambda)}-i\Gamma_{1,2}^{(\lambda)}=\frac{E^{(0)}_{1}+E^{(0)}_{2}+\lambda}{2}
\qquad
\label{E12}
\\
\pm\sqrt{\biggl(\frac{E^{(0)}_{1}\I{-}E^{(0)}_{2}}{2}\biggr)^2+\biggl(\frac{\lambda}{2}\biggr)^2+\lambda\,\frac{E^{(0)}_{1}\I{-}E^{(0)}_{2}}{2}\cos 2\vartheta}
\nonumber
\end{eqnarray} 
become degenerate for $\lambda$ equal  to 
\begin{equation}
\lambda_{{\rm EP}\pm}=-\bigl(E^{(0)}_{1}-E^{(0)}_{2}\bigr)\,e^{\pm 2 i\vartheta}
\,.
\end{equation}
These are complex conjugate EPs of the matrix \eqref{H2}, which both represent square-root singularities of the function ${{\cal E}_1^{(\lambda)}-{\cal E}_2^{(\lambda)}=\sqrt{(\lambda-\lambda_{{\rm EP}+})(\lambda-\lambda_{{\rm EP}-})}}$.
At $\lambda=\lambda_{{\rm EP}\pm}$, the Hamiltonian can undergo a similarity transformation to the Jordan form
\begin{equation}
\hat{H}^{(\lambda_{{\rm EP}\pm})}\rightarrow
\left(\begin{matrix}{\cal E}_{{\rm EP}\pm}&1\\0&{\cal E}_{{\rm EP}\pm}\end{matrix}\right)
\,,
\end{equation}
where ${{\cal E}_{{\rm EP}\pm}={\cal E}_{1}^{(\lambda_{{\rm EP}\pm})}={\cal E}_{2}^{(\lambda_{{\rm EP}\pm})}}$.
This means that two right eigenvectors $\ket{1_{\rm R}^{(\lambda)}}$ and $\ket{2_{\rm R}^{(\lambda)}}$ associated with the eigenvalues ${\cal E}_1^{(\lambda)}$ and ${\cal E}_2^{(\lambda)}$ for ${\lambda\neq\lambda_{{\rm EP}\pm}}$ contract to a single one $\ket{1_{\rm R}^{(\lambda_{{\rm EP}\pm})}}$ at the EP, and the same happens to left eigenvectors.
The unique eigenvector at the EP is self-orthogonal in the sense ${\scal{1_{\rm L}^{(\lambda_{{\rm EP}\pm})}}{1_{\rm R}^{(\lambda_{{\rm EP}\pm})}}=0}$.

Setting the real part $\epsilon$ of parameter $\lambda$ to ${\rm Re}\lambda_{{\rm EP}\pm}=\bigl(E^{(0)}_{2}\I{-}E^{(0)}_{1}\bigr)\cos 2\vartheta$ and varying the imaginary part along a half-line ${\gamma\in[0,\infty)}$, we proceed in opening the system along a trajectory passing the EP located in the lower complex half-plane of $\lambda$.
It follows from Eq.\,\eqref{E12} that for $\gamma$ increasing from zero towards the EP absolute value, the widths $\Gamma_1^{(\lambda)}$ and $\Gamma_2^{(\lambda)}$ are both equal to $\frac{1}{2}\gamma$, whereas the real energies $E_1^{(\lambda)}$ and $E_2^{(\lambda)}$ differ and collapse, according to the square-root formula, to the EP degeneracy  as $\gamma\to|{\rm Im}\lambda_{\rm EP\pm}|={|\bigl(E^{(0)}_{2}\I{-}E^{(0)}_{1}\bigr)\sin 2\vartheta|}$.
For $\gamma$ increasing further from  ${|{\rm Im}\lambda_{\rm EP\pm}|}$, the real energies stay degenerate whereas the decay widths  exhibit the square-root bifurcation to a short- and long-living asymptotic forms $\Gamma_{\rm short}^{(\lambda)}\I{=}\gamma$ and $\Gamma_{\rm long}^{(\lambda)}\I{\propto}\gamma^{-1}$.
This critical (non-analytic) form of the superradiant scenario is illustrated in Fig.\,\ref{2D} for model settings in which $\lambda_{{\rm EP}\pm}=\pm i$.

For an increasing distance  of $\epsilon$ from ${\rm Re}\lambda_{{\rm EP}\pm}$, the sharpness of the NHSR transition decreases, but its principal features remain preserved.
These cases are also illustrated in Fig.\,\ref{2D}.
We see that shorter- and longer-living states exist already at small values of $\gamma$ where their widths increase linearly.
The ratio between the larger and smaller width grows with the distance from ${\rm Re}\lambda_{{\rm EP}\pm}$. 
As $\gamma$ reaches values near or above ${\gamma\approx|{\rm Im}\lambda_{{\rm EP}\pm}|}$, one of the widths makes a crossover to the $\propto\gamma^{-1}$ behavior.
At the same time, evolution of real energies with $\gamma$ increasing across the EP region for $\epsilon\I{\neq}{\rm Re}\lambda_{{\rm EP}\pm}$ shows a smooth reduction of the spacing $|E_{1}^{(\lambda)}\I{-}E_{2}^{(\lambda)}|$.


\begin{thebibliography}{99}
\bibitem{Fes} H. Feshbach, Ann. Phys. {\bf 5}, 537 (1958); {\bf 19}, 287 (1962). 
\bibitem{Moi11} N. Moiseyev, {\it Non-Hermitian Quantum Mechanics} (Cambridge University Press, Cambridge UK, 2011).
\bibitem{Kle85} P. Kleinw{\"a}chter and I. Rotter, Phys. Rev. C {\bf 32}, 1742 (1985).
\bibitem{Sok88} V.V. Sokolov and V. Zelevinsky, Phys. Lett. B {\bf 202}, 10 (1988).
\bibitem{Sok92} V.V. Sokolov and V. Zelevinsky, Ann. Phys. {\bf 216}, 323 (1992).
\bibitem{Dic54} R.H. Dicke, Phys. Rev. {\bf 93}, 99 (1954).
\bibitem{Rot91} I. Rotter, Rep. Prog. Phys. {\bf 54}, 635 (1991).
\bibitem{Vol04} A. Volya and V. Zelevinsky, AIP Conf. Proc. {\bf 777} (2005), ed. V. Zelevinsky, p. 229.
\bibitem{Aue11} N. Auerbach and V. Zelevinsky, Rep. Prog. Phys. {\bf 74}, 106301 (2011).
\bibitem{Cel11} G.L. Celardo, N. Auerbach, F.M. Izrailev, and V.G. Zelevinsky,  Phys. Rev. Lett. {\bf 106}, 042501 (2011).
\bibitem{Liu14} C. Liu, A. Di Falco, and A. Fratalocchi, Phys. Rev. X {\bf 4}, 021048 (2014).
\bibitem{Ele14} H. Eleuch and I. Rotter, Eur. Phys. J. D {\bf 68}, 74 (2014).
\bibitem{Rot15} I. Rotter and J.P. Bird, Rep. Prog. Phys. {\bf 78}, 114001 (2015).
\bibitem{Gre15} Ya.S. Greenberg and A.A. Shtygashev, Phys. Rev. A {\bf 92}, 063835 (2015).
\bibitem{Ele17a} H. Eleuch and I. Rotter, Phys. Rev. A {\bf 95}, 022117 (2017).
\bibitem{Ele17b} H. Eleuch and I. Rotter, Phys. Rev. E {\bf 95}, 062109 (2017).
\bibitem{Jos18} S. Joshi and I. Galbraith, Phys. Rev. A {\bf 98}, 042117 (2018).
\bibitem{Kat66} T. Kato, {\it Perturbation Theory of Linear Operators} (Springer, New York, 1966).
\bibitem{Zir83} M. R. Zirnbauer, J.J.M. Verbaarschot, and H.A. Weidenm{\"u}ller, Nucl. Phys. A 411, 161 (1983).
\bibitem{Hei89} W. D. Heiss, Z. Phys. A 329, 133 (1989).
\bibitem{Cej07} P. Cejnar, S. Heinze, and M. Macek, Phys. Rev. Lett. 99, 100601 (2007).
\bibitem{Sin17} M. {\v S}indelka, L. F. Santos, and N. Moiseyev, Phys. Rev. A 95, 010103(R) (2017).
\bibitem{Str18} P. Str{\'a}nsk{\'y}, M. Dvo{\v r}{\'a}k and P. Cejnar, Phys. Rev. E {\bf 97}, 012112 (2018).
\bibitem{Hei98} W.D. Heiss, M. M{\"u}ller, and I. Rotter, Phys. Rev. E {\bf 58}, 2894 (1998).
\bibitem{Jun99} C. Jung, M. M{\"u}ller, and I. Rotter, Phys. Rev. E {\bf 60}, 114 (1999).
\bibitem{Lip65} H. J. Lipkin, N. Meshkov, and A. J. Glick, Nucl. Phys. {\bf 62}, 188 (1965).
\bibitem{Hol40} T. Holstein and H. Primakoff, Phys. Rev. 58, 1098 (1940).
\bibitem{Str19} M. Macek, P. Str{\'a}nsk{\'y}, A. Leviatan and P. Cejnar, Phys. Rev. C 99, 064323 (2019). 
\bibitem{Meh04} M.L. Mehta, {\it Random Matrices} (Academic, London, 2004). 
\bibitem{Wol80} H. Wolkowicz and G.P.H. Styan, Linear Algebra Appl. {\bf 29}, 471 (1980).
\end{thebibliography}
\end{document}